\documentclass[9pt,twocolumn,twoside]{osajnl}

\journal{jocn} 

\setboolean{shortarticle}{false}

\usepackage[T1]{fontenc}
\usepackage[utf8]{inputenc}
\usepackage{amsmath}
\usepackage{graphicx}
\usepackage{subcaption}
\usepackage[numbers,sort&compress]{natbib}
\usepackage{hyperref}
\usepackage{booktabs}
\usepackage{array}
\usepackage{tabularx}
\usepackage{makecell}
\usepackage{rotating}
\usepackage{algorithm}
\usepackage{algpseudocode}
\algrenewcommand\algorithmicindent{0.5em}%

\hypersetup{
    colorlinks=true,
    linkcolor=blue,
    citecolor=blue,
    urlcolor=blue
}

\AtBeginDocument{%
}

\title{A Wavelength Borrowing Architecture for Optical Data Center Networks - Extended Version}

\author[1]{Andrea Detti}
\author[1]{Chiara Lodovisi}
\author[1]{Silvello Betti}

\affil[1]{CNIT - University of Rome ``Tor Vergata'', Electronic Engineering Dept., Italy}

\begin{abstract}
The growth of east-west traffic, along with the cost and power consumption of electronic switching, is motivating the integration of a low-power, high-rate, all-optical layer within the data center network. This paper presents a spine-leaf all-optical architecture in which the default wavelength configuration, one wavelength per source--destination leaf pair, can be reconfigured to accommodate unbalanced traffic demand: wavelengths that are unused or lightly loaded at one leaf are borrowed by another leaf with higher demand. This topology-engineering capability is combined with a traffic-engineering scheme, based on two-hop detouring, enabling the control of wavelength load while limiting the amount of detoured traffic. A key feature of the architecture is that its degree of wavelength reconfigurability is set by a single tunable parameter, the borrowing degree $B$, ranging from none to full; performance evaluation shows that performance nearly equal to that of a fully reconfigurable architecture is achieved well below the maximum $B$, saving the complexity of a fully reconfigurable solutions. The optical fabric relies on mature, data-center-grade components, namely AWGs, AWGRs, colorless OXCs, and combiners. The reconfiguration speed and the adopted dynamic control policy make the architecture suitable for network tiers where the traffic varies on timescales of tens of minutes or more, e.g., among groups of racks (pods).
\end{abstract}

\setboolean{displaycopyright}{false} 

\begin{document}

\maketitle

\section{Introduction}

East-west traffic, from server-to-server or rack-to-rack, dominates volume in modern data centers, driven by distributed applications, storage replication, and increasingly by large-scale distributed training and inference workloads that demand high bandwidth between servers and racks~\cite{kandula2009nature}.

Spine-leaf network architectures are commonly used to support this need~\cite{alfares2008}. A leaf node is an L3 Ethernet switch: some of its ports serve either a single rack, acting as a Top-of-Rack (ToR) switch, or a group of ToR switches forming a Pod, while the remaining ports connect to a bank of spine L3 switches, providing leaf-to-leaf connectivity.

Increasing the leaf-to-leaf bandwidth requires scaling the spine-leaf segment. When feasible, this can be achieved by raising the Ethernet line rate (e.g. from 100~Gbps to 400~Gbps); such an upgrade keeps the fiber plant untouched but requires replacing switches, or at least their optical transceivers, at every leaf and spine switch. Alternatively, capacity can be added by deploying more parallel leaf-to-spine fibers or additional spine switches, but this entails rewiring and possibly a hardware upgrade whenever free ports are unavailable on the switches. 

To simplify bandwidth scaling and lower data centre network costs, cloud service providers and the academic community are exploring the possibility of replacing the electronic spine layer with an all-optical one (see ~\cite{kachris2012,baziana2024optical} for surveys). Leaf-to-leaf traffic is carried end-to-end over wavelengths routed through an optical fabric without any intermediate optoelectronic conversions~\cite{poutievski2022jupiter,sirius2020,patronas2025,baziana2024optical}. This optical spine reduces energy consumption, as neither buffering nor electronic processing is required; it halves the number of expensive transceivers, which are needed only on the leaf side, and is transparent to data format and speed; therefore, migration to a newer Ethernet generation requires only the upgrading of the leaf transceivers, whilst the optical backbone remains unchanged.

Since optical switching usually does not support buffering, a wavelength used by a leaf to receive traffic can be used at the same time by only one source leaf with no in-network resource contention. This raises an end-to-end wavelength assignment problem that adapts the wavelength provisioning to traffic demand. The wavelength assignment strategy can be regarded as a \textit{topology engineering} problem, which can nonetheless be coupled with a \textit{traffic engineering} one: how to route traffic on top of the optical topology, possibly accepting intermediate opto-electronic conversion~\cite{hyperflex}. For instance, if the topology engineering solution assigns no wavelength between leaves $j$ and $d$, traffic from $j$ to $d$ must instead be detoured through an intermediate leaf $i$, which has wavelengths towards both $j$ and $d$, thus forming a two-hop $j\to i\to d$ path \footnote{Orthogonal to both topology and traffic engineering is a third optimization axis, \textit{placement engineering}, which we do not consider in this paper. Rather than adapting the network to the traffic, it acts upstream, at the traffic-source level, e.g., through traffic-aware virtual machine placement, to reduce the load that the network must carry in the first place~\cite{meng2010vmplacement}.}.

Most spine optical fabrics proposed in the literature, however, are designed for either no reconfigurability~\cite{lightwave2023} or full wavelength-level reconfigurability, where each receiver wavelength can be assigned to any source, and/or rely on wavelength-selective switches (WSS) or in-network optical signal processing~\cite{hyperflex,baziana2024optical,rotos2020}. Optical signal processing remains laboratory-grade technology, while WSS devices are currently expensive, complex, and limited in port count, compared to simpler devices, with no optical processing, such as the colorless Optical Cross-Connect (OxC) or the Arrayed Waveguide Grating Router (AWGR)---technology already deployed in real data centers~\cite{poutievski2022jupiter,lightwave2023} or experimentally demonstrated at scale~\cite{sirius2020}. 

In this paper we propose a modular architecture for an optical spine layer that closes all three gaps: it dispenses with optical signal processing, avoiding laboratory-grade technology; it dispenses with WSS, avoiding their cost, complexity, and port-count limits; and it replaces the all-or-nothing reconfigurability choice with a reconfiguration capability, and hence system cost/complexity, that can be scaled out gradually by adding optical devices only where needed.

The baseline configuration provides a single \textit{default} wavelength between any leaf pair. To adapt wavelength assignment to traffic demand, a leaf with idle capacity toward a given destination---the \textit{donor}---can \textit{lend} its default wavelength to another leaf that needs additional capacity toward that same destination---the \textit{borrower}. Any residual traffic that the donor still has toward the destination is  split, via traffic engineering, across two-hop detours through other leaves. The complexity-reconfigurability tradeoff of the architecture is governed by a single \textit{borrowing degree} $B$: the number of donors a leaf can concurrently borrow from, and symmetrically the number of borrowers a leaf can concurrently lend to, is at most $B-1$. This parameter determines the wavelength reconfiguration capability of the architecture, along with its complexity and cost.

A fully reconfigurable architecture, where any receiver wavelength can be assigned to any source, would require $B=L$, where $L$ is the number of leaf nodes. However, we show that full reconfigurability is not necessary to achieve the desired performance objective: the borrowing architecture can instead be tuned in a cost-adaptive way.

Overall, the contributions of this paper are:
\begin{itemize}
    \item We propose an optical spine-leaf architecture whose reconfigurability level, hence  complexity, can be tuned to fit data center needs, relying only on commercially mature technology---AWGR, AWG multiplexers/demultiplexers, passive optical combiners, and a colorless OxC.
    \item We model the architecture limits as a set of optical and electronic mixed-integer linear-programming (MILP) constraints, paving the way for any related topology/traffic engineering optimizations.
    \item We focus on one optimization objective: keeping every wavelength load below a threshold with the minimum volume of detoured traffic. A greedy heuristic jointly selects the borrowing configuration (topology engineering) and the detouring fractions (traffic engineering); its core routing subroutine, two-hop water-filling (\textsc{2HWF}), can also serve as a stand-alone traffic engineering solution.
    \item We propose a dynamic control policy for time-varying traffic: it adapts the detouring fractions at every traffic update and reconfigures the optical topology only when the maximum wavelength load exceeds a tolerance, migrating via a sequence of wavelength switches that keep loads bounded and avoid traffic interruption. It targets demand persisting over tens of minutes, positioning the solution as coarse-grained circuit switching rather than packet switching.
\end{itemize}

The remainder of the paper is organized as follows. Section~\ref{sec:arch} describes the proposed wavelength-borrowing architecture in detail. Section~\ref{sec:problem} formulates the wavelength assignment and two-hop detouring problem. Section~\ref{sec:heuristic} presents the greedy heuristic and dynamic control policy. Section~\ref{sec:perf} compares the proposed architecture and topology/traffic engineering algorithm with a theoretical optimal solution and simple solutions representative of a static, non-borrowing optical core, with and without traffic engineering. Finally, Section~\ref{sec:related} discusses related work.

\section{Architecture Description}
\label{sec:arch}
\subsection{Overview}
\begin{figure}[t]
      \centering
      \includegraphics[scale=0.50]{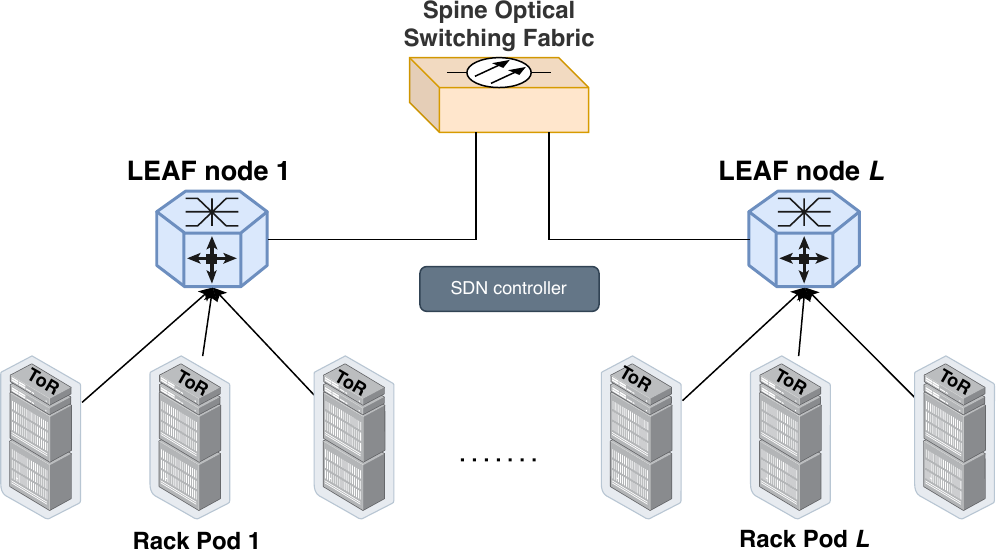}
      \caption{Architecture of the wavelength borrowing data center}
      \label{fig:osp}
\end{figure}
\begin{figure*}[t]
      \centering
      \includegraphics[scale=0.53]{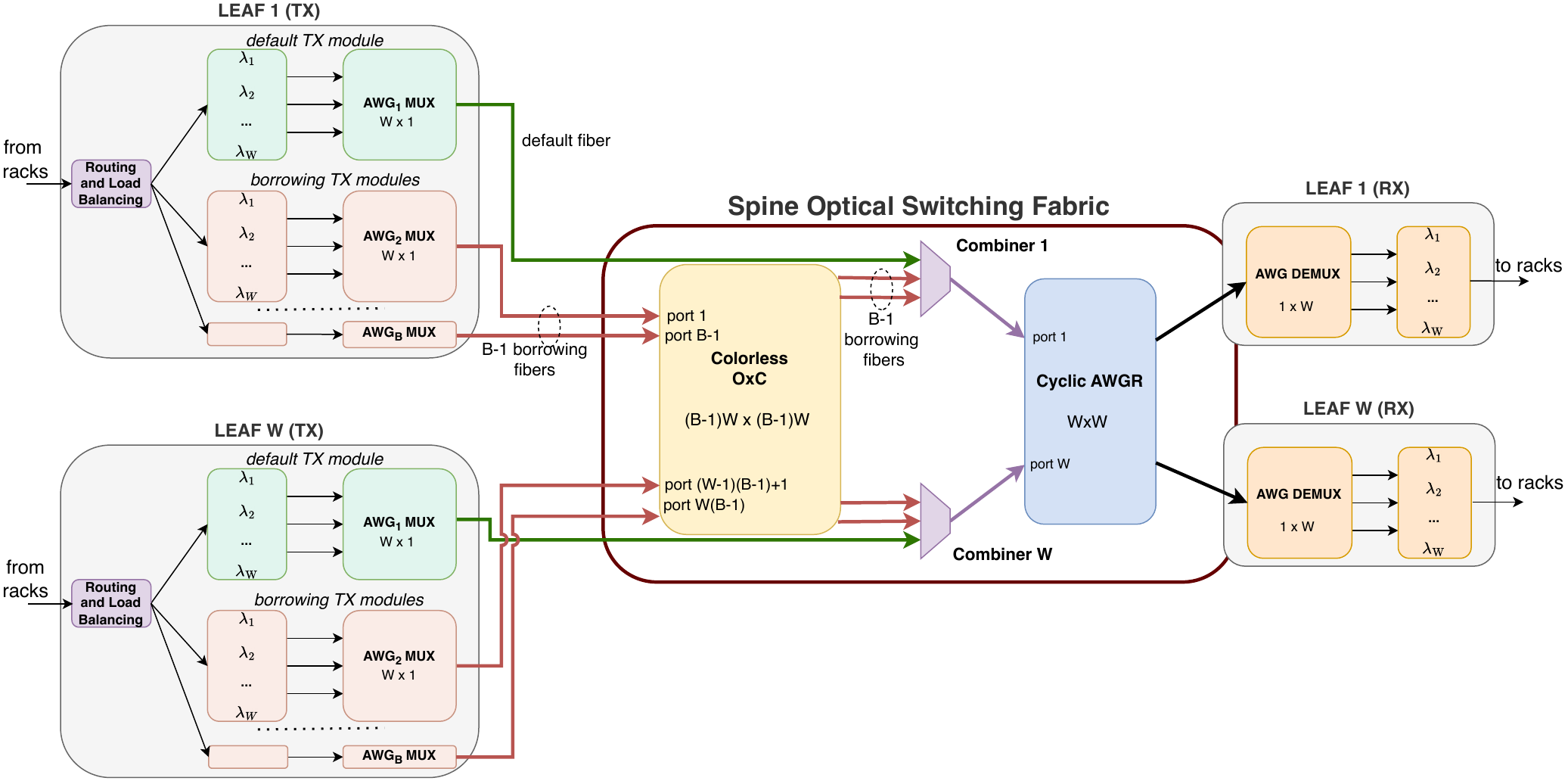}
      \caption{Schematic of the wavelength borrowing architecture.}
      \label{fig:arch}
\end{figure*}
As shown in \autoref{fig:osp}, the proposed architecture comprises $L$ rack groups/pods, each connected to a \textit{leaf} node through a Top-of-Rack (ToR) switch over standard Ethernet. Inter-leaf traffic is carried over all-optical circuits using Wavelength Division Multiplexing (WDM) with $W$ wavelengths. A \textit{spine optical switching fabric} routes these circuits, while an SDN controller manages their configuration~\cite{openroadm2024}.

Under default operation, each leaf node uses one \textit{default} wavelength per destination, yielding a fully balanced allocation of optical resources across all leaves. The key innovation is a dynamic \emph{wavelength borrowing} mechanism: a leaf with idle capacity---the \emph{donor}---lends its unused default wavelengths to other leaves---the \emph{borrowers}---increasing each borrower's instantaneous bandwidth toward a specific destination. For example, leaf~1 and leaf~$L$ each have a default wavelength, $\lambda_2$ and $\lambda_1$ respectively, to reach leaf~2. When leaf~1 has little or no traffic toward leaf~2, it can lend $\lambda_2$ to leaf~$L$, expanding the latter's available wavelengths toward leaf~2 to $\{\lambda_1, \lambda_2\}$.

The degree of reconfigurability is governed by a parameter $B$ termed the \textit{borrowing degree}. Specifically, the number of leaves from which a leaf may concurrently borrow or lend resources is at most $B-1$. Increasing $B$ improves wavelength allocation flexibility at the cost of additional optical hardware. The architecture thus offers a tunable trade-off between hardware complexity/cost and resource allocation flexibility.

\autoref{fig:arch} shows the optical components implementing the transmitting (left) and receiving (right) functionalities of the leaves, together with the interconnecting spine optical fabric. Each block is described in the following subsections, covering components and wiring first, followed by data transfer operations. For simplicity, the description focuses on the case where the number of leaves equals the number of wavelengths, i.e., $L=W$. Appendix I in \cite{extended} extends the discussion to the more general case.

\subsection{Components and Wiring}
\subsubsection{Leaf Nodes}
\paragraph*{Transmitting operations.}
For data transmission, each leaf node contains $B$ copies of a transmission module, each composed by a bank of $W$ fixed-wavelength lasers ($\lambda_1, \ldots, \lambda_W$) feeding a $W{\times}1$ AWG multiplexer whose output fiber connects to the spine optical fabric.

The first module of a leaf $i$ (green in the figure) is the \textit{default} one: its \textit{default fiber} connects directly to the $i$-th combiner, bypassing the spine OxC, and its $W$ wavelengths are the default ones toward the remote leaves, one per leaf, switched off and lent to borrowing leaves as needed.

The remaining $B-1$ \textit{borrowing modules} (light-red in the figure) are each connected to the spine OxC via a \textit{borrowing fiber} 
and transmit over one or more wavelengths borrowed from a single donor leaf; since wavelengths from different donors require different borrowing modules, at most $B-1$ donors can be used concurrently. Any laser within a borrowing module can be activated on demand, according to the wavelength assignment strategy implemented by the SDN controller.

Finally, an ingress SDN-controlled load balancer routes outgoing traffic to the buffers drained by the lasers of the transmission modules, following a specific traffic engineering strategy.

\paragraph*{Reception operations.}
For data reception, each leaf node is connected to the spine optical fabric via a single input fiber carrying $W$ wavelengths. An AWG demultiplexer separates each wavelength onto a dedicated fiber; the transported bit stream is then converted to the electronic domain by a dedicated WDM receiver for subsequent packet forwarding, either to the final rack or to the next-hop leaf in case of detouring.

\subsubsection{Spine Optical Switching Fabric}

The spine optical switching fabric consists of three elements: a colorless Optical Cross-Connect (OxC), a bank of combiners, and a cyclic Arrayed Waveguide Grating Router (AWGR).

\paragraph*{Cyclic AWGR.}
The AWGR is a fully passive component that routes each wavelength $\lambda_n$ arriving at input port $i$ to a deterministic output port $d$ according to the cyclic routing rule:
\begin{equation}
    d = (n - i) \bmod W + 1,
    \label{eq:awgr_routing}
\end{equation}
Hence, the $W$ wavelengths entering any input port are routed to $W$ distinct output ports, and each output port receives every wavelength exactly once, from different input ports, so that signals sharing a wavelength never collide.

Each AWGR output port is connected to a specific destination leaf: output port $d$ is connected to leaf $d$, and each AWGR input port is connected to a dedicated combiner.

The AWGR can be realized as a single device or replaced by Sato's cascaded small cyclic AWG architecture~\cite{sato2018}, which synthesizes a $KM{\times}KM$ equivalent switch from $M$ copies of $K{\times}K$ AWGs and $K$ copies of $M{\times}M$ AWGs, with $K$ and $M$ mutually coprime integers. A multi-stage Thin-CLOS wavelength-routing fabric built from smaller AWGRs offers an alternative, experimentally demonstrated scale-out solution ~\cite{proietti2018thinclos}.

\paragraph*{Combining stage.}
Combiner $i$, connected to AWGR port $i$, merges optical signals from $B$ fibers: i) the default fiber from $\text{AWG}_1$ of node $i$, and ii) a group of $B-1$ fibers arriving from the OxC, each carrying the wavelengths of a distinct borrowing fiber.
The combiner size $B$ limits a donor node to serving at most $B-1$ borrowers simultaneously\footnote{The number of borrowing TX modules per node and the number of borrowing input fibers per combiner are both equal to $B-1$. Relaxing this equality by introducing two distinct parameters---$B_{\text{TX}}$ for the TX module count and $B_{\text{COMB}}$ for the combiner size---yields an asymmetric resource relocation constraint: a borrower may use at most $B_{\text{TX}}-1$ donors, while a donor may lend resources to at most $B_{\text{COMB}}-1$ borrowers.}.

The wavelength sets carried by the $B$ input fibers of any combiner are \emph{guaranteed to be disjoint} by the wavelength assignment strategy. Accordingly, the combiner can be implemented as a passive coupler, resulting in a simple, fully passive design, but incurring an optical loss of $10\log_{10}(B)$~dB, which could be compensated by a optical amplifier at the combiner output.

\paragraph*{Colorless OxC.}
The OxC has size $(B-1)W \times (B-1)W$ and routes the wavelengths of borrowing fibers to combiners as configured by the SDN controller. The OxC performs purely spatial switching with no wavelength awareness. Currently, MEMS technology is a valuable choice for the OxC implementation, as it can achieve the required port count (e.g., on the order of hundreds) with acceptable insertion loss and switching time~\cite{lightwave2023}.

\subsection{Leaf-to-Leaf Data Transfer}
\begin{figure}[t]
      \centering
      \includegraphics[scale=0.6]{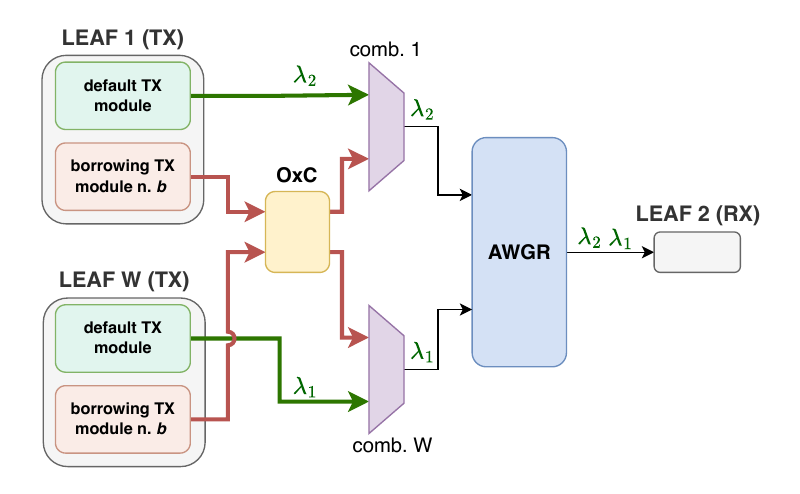}
      \caption{Fully balanced configuration, no borrowing. Leaf 1 and leaf $W$ reach leaf 2 with their default wavelengths $\lambda_2$ and $\lambda_1$, respectively.}
      \label{fig:balanced}
\end{figure}
\begin{figure}[t]
      \centering
      \includegraphics[scale=0.55]{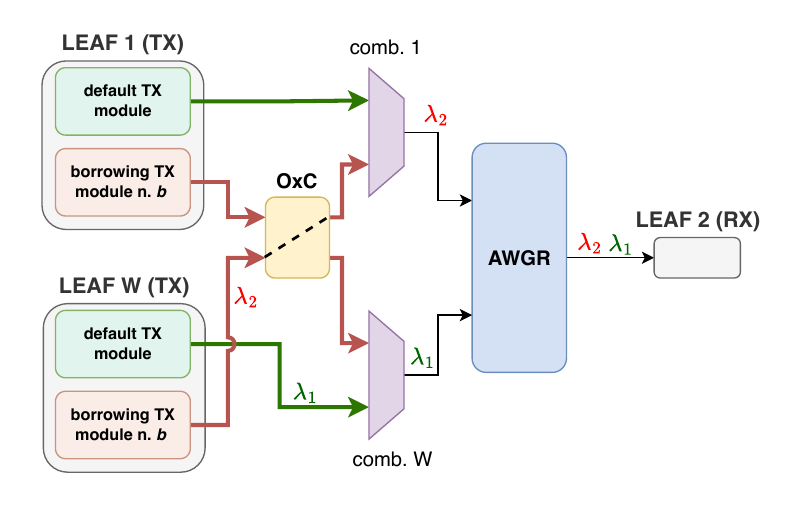}
      \caption{Unbalanced configuration with wavelength borrowing. Leaf 1 lends the default wavelength $\lambda_2$ to leaf $W$. The resulting optical capacity from leaf $W$ to leaf 2 comprises  $\lambda_2$ and $\lambda_1$.}
      \label{fig:unbalanced}
\end{figure}

\subsubsection{Fully-Balanced Configuration}
Under a fully-balanced traffic pattern, no wavelength borrowing takes place, and each leaf uses only its default TX module to simultaneously reach all $W$ destinations, one default wavelength per destination. For instance, in \autoref{fig:balanced}, leaf~1 and leaf~$W$ use their default wavelengths $\lambda_2$ and $\lambda_1$ to reach leaf~2, respectively

\subsubsection{Unbalanced Configuration}
During an unbalanced traffic configuration, a leaf $i$ can require extra bandwidth toward a destination $d$, while another leaf $j$ has its default wavelength toward $d$ idle or lightly loaded. Two approaches can handle this traffic variation. The first is an electronic-only approach based on \textit{traffic detouring}~\cite{poutievski2022jupiter}: the excess traffic from the overloaded leaf $i$ toward $d$ is rerouted over a two-hop path through leaf $j$ (or more than one), processed electronically there, and then forwarded on that leaf's unloaded default wavelength to $d$. The second is an optical--electronic hybrid approach based on wavelength borrowing and detouring: the underloaded leaf $j$ lends its default wavelength toward $d$ to $i$; the borrower leaf $i$ activates the laser of the borrowed wavelength on a borrowing TX module dedicated to wavelengths borrowed from donor $j$, and the OxC routes the corresponding borrowing fiber to the combiner of donor leaf $j$. Any residual traffic from donor leaf $j$ toward $d$ is detoured through other leaves that still have an active default wavelength to $d$.

For instance, in \autoref{fig:unbalanced}, leaf~1 lends its default wavelength $\lambda_2$ to leaf~$W$. Leaf~$W$ uses one of its borrowing TX modules to transmit data on the borrowed wavelength $\lambda_2$. The OxC routes $\lambda_2$ from the borrowing TX module of leaf~$W$ to the donor combiner~1, and the AWGR then routes $\lambda_2$ to leaf~2. Consequently, leaf~$W$ can use two wavelengths to reach destination leaf~2.


\section{Wavelength Assignment and Traffic Detouring}
\label{sec:problem}
The wavelength borrowing architecture can be dynamically controlled to achieve different optimization goals. In this paper, we consider the base architecture in \autoref{fig:arch} with a number of leaves equal to the number of wavelengths, i.e., $L=W$. The architectural extensions in Appendix I in \cite{extended} are left for future work.
The following subsections first derive the problem constraints defining the feasible region for any optimization within the wavelength borrowing framework with two-hop detouring, and then present our specific optimization problem.


\subsection{Variables and Constraints}
Let $A^E_{i,d}$ denote the \textit{end-to-end} traffic generated by racks served by leaf $i$ and directed to racks of leaf $d$, normalized to the bitrate of one wavelength, i.e., $A^E_{i,d} = 1$ means a traffic bitrate equal to the wavelength one; we collect these entries into the traffic matrix $\mathbf{A}^E = \{A^E_{i,d}\}$. The borrowing configuration is represented by the binary matrix $\mathbf{b} = \{b_{i,j,d}\}$, where an entry equal to 1 indicates that leaf $i$ borrows the default wavelength used by leaf $j$ to reach destination leaf $d$. The traffic detouring configuration is represented by the matrix $\mathbf{w} = \{w_{j,i,d}\}$, whose entries denote the fraction of end-to-end traffic $A^E_{j,d}$ that is electronically detoured through node $i$.

\paragraph{Wavelength assignment constraints.}
We define the following variables related to the optical architecture in \autoref{fig:arch}:
\begin{align}
    &x_{i,j} = \min\!\left(1,\sum_d b_{i,j,d}\right) \\
    &d_{i,d} = 1 - \sum_j b_{j,i,d} \\
    &c_{i,d} = d_{i,d} + \sum_j b_{i,j,d} \label{eq:c}
\end{align}
where $x_{i,j}$ is a binary variable indicating that leaf $i$ borrows at least one wavelength from leaf $j$; $d_{i,d}$ indicates that leaf $i$ retains its default wavelength towards destination leaf $d$; and $c_{i,d}$ denotes the total number of wavelengths available from leaf $i$ to destination leaf $d$, collected into the capacity matrix $\mathbf{c} = \{c_{i,d}\}$.

The hardware limit of the architecture imposes that a wavelength assignment resulting from the borrowing configuration $\mathbf{b}$ is feasible only if it satisfies the following constraints:
\begin{align}
    & \sum_i x_{i,j} < B \quad \forall j \label{eq:optc1}\\
    & \sum_j x_{i,j} < B \quad \forall i \label{eq:optc2}\\
    & \sum_{i} b_{i,j,d} \leq 1 \quad \forall j,d \label{eq:optc3} \\
    & \sum_k b_{j,k,d} = 0  \quad \forall j,d : \sum_{i} b_{i,j,d} = 1 \label{eq:optc4} \\
    & r_{i,j} = \min\!\left(1, c_{i,j}\right) , \quad r_{i,d} + \sum_j r_{i,j} \; r_{j,d} > 0 \quad \forall i,d : i \neq d
    \label{eq:optc5}
   \end{align}
The constraints~\eqref{eq:optc1}--\eqref{eq:optc2} reflect the limit of $B-1$ TX modules and combiner ports available per leaf for wavelength borrowing; constraint~\eqref{eq:optc3} ensures that the default wavelength of leaf $j$ towards destination $d$ is borrowed by at most one leaf; constraint~\eqref{eq:optc4} ensures that a leaf $j$ lending its default wavelength towards destination $d$ may not simultaneously borrow any wavelength towards the same destination. Finally, $r_{i,j}$ indicates whether leaf $i$ has at least one wavelength (default or borrowed) towards leaf $j$, and \eqref{eq:optc5} enforces that every ordered source--destination \textit{pair} of distinct leaves $(i,d)$ remains connected within at most two hops: either $i$ has a direct wavelength to $d$ ($r_{i,d}=1$), or there exists at least one intermediate leaf $j$ with $r_{i,j}=1$ and $r_{j,d}=1$, thereby guaranteeing full leaf-to-leaf connectivity in at most two hops.

\paragraph{Traffic detouring constraints.}
A traffic detouring solution $\mathbf{w}$ is subject to the following constraints:
\begin{align}
    & \sum_i w_{j,i,d} \leq 1 \quad \forall j,d \label{eq:elc1}\\
    & w_{j,i,d} \geq 0 \quad \forall j,i,d \label{eq:elc2}\\
    & w_{j,i,d} = 0 \quad \forall j,d,i : c_{i,d}=0 \; \text{or} \; c_{j,i}=0\label{eq:elc3} \\
    & A^E_{j,d} = A^{H0}_{j,d} + A^D_{j,d} \quad \forall j,d \label{eq:elc4}
\end{align}
Constraint~\eqref{eq:elc1} ensures that the total detoured fraction of any source's traffic does not exceed unity; constraint~\eqref{eq:elc2} requires non-negative detouring fractions; constraint~\eqref{eq:elc3} restricts detoured traffic to be forwarded only through leaves with available two-hop connectivity; and constraint~\eqref{eq:elc4} is the  traffic conservation condition for any pair $(j,d)$, requiring that the non-detoured (direct) traffic $A^{H0}_{j,d}$ and the detoured traffic $A^D_{j,d}$ together equal the total end-to-end traffic $A^E_{j,d}$.

\subsection{Optimization objective}

We formulate a single objective: minimize the total electronically detoured traffic $T^D$ while ensuring that the traffic load $\rho_{i,d}$ of wavelengths between any source--destination pair $(i,d)$ is below a given threshold $\rho_\text{th}$. For instance, $\rho_\text{th} = 0.9$ implies that the average traffic offered to wavelengths of any pair $(i,d)$ is lower than 90\% of the maximum wavelengths' bitrate.

The optimization thus determines the borrowing configuration $\mathbf{b}$, which is binary, and the detouring fractions $\mathbf{w}$, which are real-valued, minimizing $T^D$ under the wavelength assignment and detouring constraints, together with the load cap condition~\eqref{eq:objc1}. The resulting problem is an MILP.
\begin{align}
    &\min_{\mathbf{b},\,\mathbf{w}}\quad
    T^D = \sum_{j,i,d} w_{j,i,d}\,A^E_{j,d}
    \label{eq:objective}\\
    &\text{s.t.} \notag\\
    &\quad \eqref{eq:optc1}\text{--}\eqref{eq:optc5},\ \eqref{eq:elc1}\text{--}\eqref{eq:elc4},\notag\\
    &\quad \rho_{i,d} \leq \rho_\text{th} \quad \forall i,d \label{eq:objc1} 
\end{align}

The load $\rho_{i,d}$ is the ratio of offered traffic to wavelength capacity of a pair $(i,d)$ and can be computed as follows. The $c_{i,d}$ wavelengths between a pair $(i,d)$ support three types of traffic:
\begin{itemize}
    \item \textit{direct} $A^{H0}_{i,d}$: the portion of end-to-end traffic $A^E_{i,d}$ forwarded by leaf $i$ to destination $d$ without detouring;
    \item \textit{local-detoured} $A^{H1}_{i,d}$: the portion of end-to-end traffic from $i$ to any destination $j \neq d$ detoured via $d$ (first-hop detouring);
    \item \textit{remote-detoured} $A^{H2}_{i,d}$: the portion of traffic from any leaf $j \neq i$ detoured via $i$ to reach leaf $d$ (second-hop detouring). This traffic is subject to a possible packet loss rate $P_{j,i}$ on the $(j,i)$ wavelengths (first-hop loss)\footnote{For the loss rate $P$, we consider a fluidic model in which the loss volume is simply equal to the amount of traffic exceeding the optical capacity $c_{i,d}$}.
\end{itemize} 

\begin{align}
    \label{eq:traffic}
    & A^{H0}_{i,d} = \left(1-\sum_j w_{i,j,d}\right)\, A^E_{i,d} \\
    & A^{H1}_{i,d} = \sum_j w_{i,d,j}\, A^E_{i,j} \\
    & A^{H2}_{i,d} = \sum_j w_{j,i,d}\, A^E_{j,d}\, \left(1-P_{j,i}\right) \\
    & P_{i,d} = \frac{\max\left(A^{H0}_{i,d} + A^{H1}_{i,d} + A^{H2}_{i,d}-c_{i,d},0\right)}{A^{H0}_{i,d} + A^{H1}_{i,d} + A^{H2}_{i,d}}
\end{align}

The resulting load on source--destination pair $(i,d)$ is:
\begin{align}
    &\rho_{i,d} = \frac{A^{H0}_{i,d} + A^{H1}_{i,d} + A^{H2}_{i,d}}{c_{i,d}}, \qquad \forall {i,d} : c_{i,d} \ge 1, \\
    &\rho_{i,d} = 0,  \qquad \forall {i,d} : c_{i,d}=0.
    \label{eq:rho}
\end{align}
These entries are collected into the load matrix $\boldsymbol{\rho} = \{\rho_{i,d}\}$.
The problem can be reduced to an MILP by introducing auxiliary binary variables that linearize the $\min(\cdot)$ terms and the product of binary variables in \eqref{eq:optc5}, and by imposing the load cap \eqref{eq:objc1} in the equivalent linear form $A^{H0}_{i,d}+A^{H1}_{i,d}+A^{H2}_{i,d}\le\rho_\text{th}\,c_{i,d}$. Since $\rho_\text{th}\le 1$, a feasible solution has no packet loss, so the loss terms $P_{i,d}$ vanish from the formulation. We solved the resulting MILP with the Gurobi optimizer, as reported in \autoref{sec:perf}.

\section{Heuristic Resource Allocation}
\label{sec:heuristic}

The joint optimization problem formulated in Section~\ref{sec:problem} couples a combinatorial selection of the borrowing variables $\mathbf{b}$ with a continuous allocation of the detouring fractions $\mathbf{w}$, and is therefore NP-hard; exact methods become computationally intractable for network sizes of practical interest. 
For this reason we developed a heuristic algorithm that decomposes the problem into three phases, each of which relies on the same traffic engineering algorithm, called \textit{two-hop water-filling} (2HWF), described first.



\subsection{Two-hop water filling}
Water-filling is a well-known algorithm that distributes an amount of ``water'' among a set of connected ``recipients'', minimizing the maximum final level of water among the recipients \cite{WF}. We used a variation of this policy to evaluate the best detouring fractions $\mathbf{w}$ for a fixed borrowing $\mathbf{b}$ configuration and resulting capacities $\mathbf{c}$.

In our context, the water amount is the traffic $A^D_{j,d}$ to detour from $j$ to $d$, i.e., the end-to-end traffic $A^E_{j,d}$ left after subtracting the direct traffic $A^{H0}_{j,d}$. Coherently with our optimization objective, the direct traffic is the maximum portion of $A^E_{j,d}$ that keeps the pair's load within $\rho_\text{th}$, thus maximizing traffic served directly -- and hence minimizing traffic to detour -- while respecting the load constraint. Specifically,

\begin{align}
  &A^D_{j,d} = A^E_{j,d} - A^{H0}_{j,d}, \quad A^{H0}_{j,d} = \min\left(A^E_{j,d} \, , \, \rho_{\text{th}} \; c_{j,d}\right) \label{eq:AD2}
\end{align}

The recipients are the set of possible two-hop paths $(j,i,d)$ from $j$ to $d$, plus the direct link $j\rightarrow d$ itself, treated as the degenerate path $i=j$ (``self detouring''). We define the \textit{normalized} level of water contained in a two-hop path $j \rightarrow i \rightarrow d$ as the \textit{two-hop load} $\rho_{j,i,d}$ with:
\begin{equation}
    \rho_{j,i,d}=\max(\rho_{j,i},\rho_{i,d})
\end{equation} 

For self-detouring, both hops coincide with the direct link, so we set $\rho_{j,j}\equiv\rho_{j,d}$ and $c_{j,j}\equiv c_{j,d}$. 

The two-hop water-filling algorithm (\textsc{2HWF}) computes the entire detouring matrix $\mathbf{w}$ by sequentially distributing, for each pair $(j,d)$, the traffic to detour $A^D_{j,d}$ across the available two-hop paths, minimizing the resulting increase in the maximum two-hop load and thereby keeping the system as far as possible from the load constraint in \eqref{eq:objc1}.

\begin{figure}
      \centering
      \includegraphics[scale=0.7]{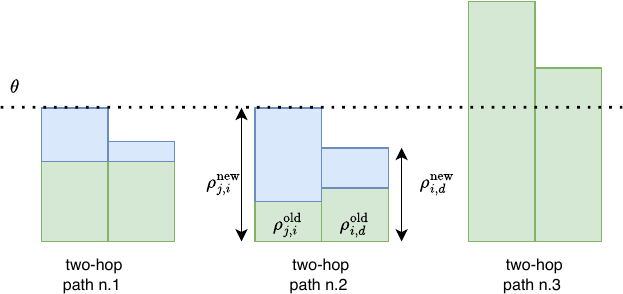}
      \caption{Two-hop water filling}
      \label{fig:wf}
\end{figure}

\autoref{fig:wf} illustrates the underlying idea of the algorithm for a single pair $(j,d)$ with three possible two-hop paths $(j,i,d)$. For each path, the two boxes show the load $\rho_{j,i}$ and $\rho_{i,d}$ of first and second hop before and after the injection of detoured traffic, marked as ``old'' and ``new'', respectively.
Starting from the old (pre-detouring) load of each pair, the algorithm searches for the minimum new two-hop load level $\theta$ such that raising the paths' two-hop loads to $\theta$ absorbs exactly the traffic to detour $A^D_{j,d}$. In the figure, the solution detours traffic only along paths 1 and 2: path 3 is already more loaded than $\theta$, so the water cannot fill that recipient\footnote{Because pairs can have a different number of wavelengths, the same traffic amount can impact their load $\rho$ differently, as for the two links of the same path in the figure.}.

Formally, for a pair $(j,d)$ for which $A^D_{j,d} > 0$, \eqref{eq:R} defines the set $R_{j,d}$ of intermediate leaves $i$ providing two-hop connectivity \mbox{$j \rightarrow i \rightarrow d$}, including self-detouring $i=j$ when the direct link exists. Using a level $\theta$, the path $(j,i,d)$ absorbs an amount of traffic $F_{\theta,j,i,d}$, given by \eqref{eq:Fs}, and the two-hop paths in $R_{j,d}$ together absorb a total traffic $F_{\theta,j,d}$, given by the first expression in \eqref{eq:F} (see Appendix II in \cite{extended}). The final level $\theta$ is then found by solving $F_{\theta,j,d} = A^D_{j,d}$, as in \eqref{eq:L}, after which the detouring fraction $w_{j,i,d}$ follows directly from the second expression in \eqref{eq:F} for $i\neq j$ and $w_{j,j,d}=0$.
\begin{align}
  & R_{j,d} = \left\{i \neq d : \left(i \neq j \wedge c_{j,i} > 0 \wedge c_{i,d} > 0\right) \vee \left(i = j \wedge c_{j,d} > 0\right)\right\} \label{eq:R} \\
  & F_{\theta,j,i,d} =
  \min\left(\max\left(0,(\theta-\rho^\text{old}_{j,i})c_{j,i}\right), \max\left(0,(\theta-\rho^\text{old}_{i,d})c_{i,d}\right)\right) \label{eq:Fs} \\
  & F_{\theta,j,d} = \sum_{i \in R_{j,d}} F_{\theta,j,i,d}, \qquad w_{j,i,d} = \frac{F_{\theta,j,i,d}}{A^E_{j,d}} \quad\forall i\neq j \label{eq:F} \\
  & \theta : F_{\theta,j,d} = A^D_{j,d}  \label{eq:L}
\end{align}

Algorithm~\ref{alg:wf} summarizes the overall procedure to distribute the whole traffic to detour, i.e., for every pair $(j,d)$ with $A^D_{j,d}>0$. The algorithm also returns the \textit{maximum overload} parameter $\Delta\rho_{\max}$, representing the maximum difference greater than zero between any load $\rho_{j,d}$ and the load threshold $\rho_\text{th}$.


\begin{algorithm}[t]
\caption{Two-Hops Water-Filling (\textsc{2HWF}) Detouring}
\label{alg:wf}
\begin{algorithmic}[1]
\Require $\mathbf{b}$ (or capacities $\mathbf{c}$), $\mathbf{A}^E$, $\rho_\text{th}$
\Ensure $\mathbf{w}$, $\boldsymbol{\rho}$, $\Delta\rho_{\max}$, $T^D$
\State Compute $c_{i,d}$ from $\mathbf{b}$ with \eqref{eq:c} $\forall i,d$, unless $\mathbf{c}$ is given 
\State Compute $A^D_{j,d}$ and $A^{H0}_{j,d}$ with \eqref{eq:AD2} $\forall j,d$
\State $\rho^\text{old}_{j,d} = A^{H0}_{j,d}/c_{j,d}$ if $c_{j,d}>0$, else $0$ $\forall j,d$
\State $w_{j,i,d} = 0$ $\forall j,i,d$
\State $S = \{(j,d) : A^D_{j,d} > 0\}$ \Comment leaf pairs with detouring traffic
\State Sort $S$ by $A^D_{j,d}$ in decreasing order
\ForAll{$(j,d) \in S$, in sorted order}
    \State Compute $R_{j,d}$ from \eqref{eq:R}
    \If{$R_{j,d} == \emptyset$} \textbf{continue} \EndIf
    \State Find water level $\theta$ solving \eqref{eq:L}
    \ForAll{$i \in R_{j,d}$}
        \State Compute $F_{\theta,j,i,d}$ from \eqref{eq:Fs}
        \If{$i \neq j$}
            \State $\rho_{j,i}^\text{new} = (\rho^\text{old}_{j,i} \; c_{j,i} + F_{\theta,j,i,d})/c_{j,i}$ \Comment{new first-hop load}
            \State $w_{j,i,d} = F_{\theta,j,i,d} / A^E_{j,d}$ \quad,\quad $\rho^\text{old}_{j,i} = \rho^\text{new}_{j,i}$
        \EndIf
        \State $\rho_{i,d}^\text{new} = (\rho_{i,d}^\text{old} \; c_{i,d} + F_{\theta,j,i,d})/c_{i,d}$  \Comment{new second-hop load}
        \State $\rho^\text{old}_{i,d} = \rho^\text{new}_{i,d}$
    \EndFor
\EndFor
\State $\Delta\rho_{\max} = \max(0,\max_{i,d}(\rho^\text{old}_{i,d}-\rho_\text{th}))$ 
\State $T^D = \sum_{j,i,d} w_{j,i,d}A^E_{j,d}$
\State \Return $\mathbf{w}$, $\boldsymbol{\rho}=\boldsymbol{\rho}^\text{old}$, $\Delta\rho_{\max}$, $T^D$
\end{algorithmic}
\end{algorithm}

\subsection{Greedy wavelength assignment and traffic detouring}
Algorithm~\ref{alg:greedy} presents the whole heuristic algorithm we use to compute the borrowing $\mathbf{b}$ and detouring $\mathbf{w}$ configurations. The algorithm is organized in three phases.

\paragraph{Phase 1: Initial water filling}
Starting from the no-borrowing state $b_{i,j,d}=0$, the algorithm computes the initial detouring fractions $\mathbf{w}$ and the corresponding load matrix $\boldsymbol{\rho}$, maximum overload $\Delta\rho_{\max}$ and $T^D$ via \textsc{2HWF}.

\paragraph{Phase 2: Greedy borrowing}
The algorithm then iterates as follows. At each round, it identifies candidate borrowing triples $(i,j,d)$ such that: there is traffic to detour on the pair $(i,d)$;
leaf $i$ is not already a donor towards $d$ ($c_{i,d} \ge 1$); and leaf $j$ retains only its default wavelength towards $d$ ($c_{j,d} == 1$) and can therefore lend it. Candidates are ranked by the score $A^D_{i,d} - A^{H0}_{j,d}$ -- the traffic currently detoured by $i$ towards $d$ minus the direct traffic that leaf $j$ would have to detour after lending its default wavelength towards $d$ -- which estimates the maximum achievable reduction in detouring traffic.

Candidates are then tested in ranked order. To limit complexity whilst accepting that the solution may not be optimal, if activating $b_{i,j,d}=1$ would violate any of the optical constraints~\eqref{eq:optc1}--\eqref{eq:optc5}, the candidate is discarded and permanently excluded from future rounds as inserted in the skip set $K$. Otherwise, the activation is applied tentatively, and the resulting values $\mathbf{w}'$, $\Delta\rho'_{\max}$, and $T'^D$ are computed by \textsc{2HWF}. The borrowing $b_{i,j,d}=1$ is confirmed, and the round restarts from the first step, if it strictly reduces $\Delta\rho_{\max}$, or leaves $\Delta\rho_{\max}$ unchanged while strictly reducing $T^D$. Otherwise, the triple is marked as rejected and permanently excluded from future rounds, and the next candidate in the ranking is tried. This acceptance rule first drives $\Delta\rho_{\max}$ toward zero, thereby satisfying the load cap in ~\eqref{eq:objc1}, and then reduces the detoured traffic volume $T^D$.
The phase terminates when a round examines every possible borrowing candidate without any useful update of the borrowing state.

\paragraph{Phase 3: Self-wavelength refinement}
Since $A^E_{j,j}=0$, a leaf $j$'s self-directed default wavelength never carries traffic, so leaving it unborrowed ($b_{i,j,j}=0\ \forall i$) after the greedy borrowing phase wastes capacity. For every such leaf $j$, the algorithm assigns this idle wavelength to the feasible leaf $i^\star$ (w.r.t.~\eqref{eq:optc1}--\eqref{eq:optc5}) with the highest current load $\rho_{i,j}$, and recomputes $\mathbf{w}$ via \textsc{2HWF}.

\paragraph{Complexity}
A single execution of \textsc{2HWF} has complexity $O(W^3\log W)$. The loop at line~7 iterates over $O(W^2)$ pairs $(j,d)$, and each iteration costs $O(W\log W)$, which is dominated by the solution of \eqref{eq:L}: since $F_{\theta,j,d}$ in \eqref{eq:F} is piecewise linear in $\theta$ with at most $2W$ breakpoints, the solution is found by sorting the breakpoints and performing a single linear sweep over them.

Regarding Algorithm~\ref{alg:greedy}, Phase~1 runs \textsc{2HWF} once, and Phase~3 costs $O(W^4\log W)$, as it runs \textsc{2HWF} $O(W)$ times. The cost of Phase~2, which dominates, is $O(E\,W^3\log W)$, where $E$ is the number of tentative \textsc{2HWF} evaluations. Since each triple $(i,j,d)$ is either applied or inserted in the skip set $K$, and is never reconsidered, $E=O(W^3)$ and the worst-case complexity of Algorithm~\ref{alg:greedy} is $O(W^6\log W)$. In practice, $E\ll W^3$, as only triples with traffic to detour are candidates, so the actual running time is much lower. The detailed derivation and measured values of $E$ are reported in Appendix~IV of \cite{extended}.

\begin{algorithm}[t]
\caption{Greedy borrowing and 2HWF detouring}
\label{alg:greedy}
\begin{algorithmic}[1]
\Require $\mathbf{A}^E$, $B$, $\rho_\text{th}$
\Ensure $\mathbf{b}$, $\mathbf{w}$, $\mathcal{H}$
\State $b_{i,j,d} = 0 \;\; \forall i,j,d$; $\mathcal{H} = ()$ \Comment{$\mathcal{H}$: borrowings sequence}
\State $\mathbf{w}, \boldsymbol{\rho}, \Delta\rho_{\max}, T^D \gets \textsc{2HWF}$ with $\mathbf{b}$ \Comment{P1: initial water filling}
\State $K = \emptyset$ \Comment{Set of $(i,j,d)$ triples to skip}
\Repeat \Comment{P2: greedy borrowing}
    \State $\text{updated} = \textbf{false}$
    \State build borrowing candidates $\mathcal{C} = \{(i,j,d): A^D_{i,d} > 0,\ c_{i,d}>0, c_{j,d}==1, (i,j,d) \notin K$\}
    \State sort $\mathcal{C}$ by score $A^D_{i,d} - A^{H0}_{j,d}$ in decreasing order
    \ForAll{$(i,j,d) \in \mathcal{C}$, in sorted order}
        \If{activating $b_{i,j,d}=1$ violates \eqref{eq:optc1}--\eqref{eq:optc5}}
            \State $K = K \cup (i,j,d)$; \textbf{continue}
        \EndIf
        \State $\mathbf{b}' = \mathbf{b}$ with $b'_{i,j,d}=1$
        \State $\mathbf{w}', \boldsymbol{\rho}', \Delta\rho'_{\max}, T'^D \gets$ \textsc{2HWF} with $\mathbf{b}'$
        \If{$\Delta\rho'_{\max} < \Delta\rho_{\max}$ \textbf{or} ($\Delta\rho'_{\max} == \Delta\rho_{\max}$ \textbf{and} $T'^D < T^D$)}
            \State $\mathbf{b} = \mathbf{b}'$; $\mathbf{w} = \mathbf{w}'$; $\Delta\rho_{\max} = \Delta\rho'_{\max}$; $T^D = T'^D$
            \State  append $(i,j,d)$ to $\mathcal{H}$; $\text{updated} = \textbf{true}$; \textbf{break}
        \Else
            \State $K = K \cup (i,j,d)$
        \EndIf
    \EndFor
\Until{\textbf{not} updated}
\ForAll{$j$ such that $b_{i,j,j}==0 \;\forall i$} \Comment{P3: refinement}
    \State $\mathcal{F}_j \gets \{i \neq j : \text{activating } b_{i,j,j}=1 \text{ satisfies \eqref{eq:optc1}--\eqref{eq:optc5}}\}$
    \If{$\mathcal{F}_j \neq \emptyset$}
        \State $i^\star \gets \arg\max_{i \in \mathcal{F}_j} \rho_{i,j}$
        \State $b_{i^\star,j,j} \gets 1$; append $(i^\star,j,j)$ to $\mathcal{H}$
        \State $\mathbf{w}, \boldsymbol{\rho}, \Delta\rho_{\max}, T^D \gets$ \textsc{2HWF} with $\mathbf{b}$
    \EndIf
\EndFor

\State \Return $\mathbf{b}$, $\mathbf{w}$, $\mathcal{H}$
\end{algorithmic}
\end{algorithm}

\subsection{Dynamic control policy}
\label{sec:dynamic}
Algorithm~\ref{alg:greedy} computes its results for a given traffic matrix $\mathbf{A}^E$, and thus applies to a static traffic regime. This section presents an SDN control policy based on it that adapts the borrowing and detouring state to time-varying traffic, under the additional constraints of a dynamic regime. In production networks, pod-to-pod demand exhibits short-term fluctuations around a long-term trend, and only long-term variations, over weeks, require topology adjustments~\cite{poutievski2022jupiter}. The policy follows this approach: traffic engineering through \textsc{2HWF} alone absorbs small traffic variations, and topology engineering via Algorithm~\ref{alg:greedy} runs only when the wavelength load approaches a critical level and topology engineering becomes necessary. However, the planned configuration cannot be applied as is, since i) some transitions could temporarily interrupt traffic, and ii) the number of new optical paths must be limited to bound the migration time.

\paragraph{Control loop.}
The policy control loop provides that, at every time window $k$, a new traffic matrix $\mathbf{A}^E(k)$ is measured. The controller recomputes only $\mathbf{w}$ with \textsc{2HWF}. Only if the resulting maximum pair load exceeds the threshold by a tolerance factor, $\rho_{\max}>\tau\rho_\text{th}$ with $\tau>1$, is the optical layer reconfigured, in three stages.
In the \emph{plan} stage, Algorithm~\ref{alg:greedy} is run from scratch on $\mathbf{A}^E(k)$ to compute a configuration $\mathbf{b}^\text{plan}$, which is not applied as it is. In the \emph{reduction} stage, the last borrowings added by Algorithm~\ref{alg:greedy} are reverted, until the number of lasers to activate $N_\text{act}$, i.e., new borrowings plus default wavelengths regained by lenders (formally defined in Appendix VI of \cite{extended}), is at most a threshold $V$, giving the target configuration $\mathbf{b}^\text{target}$. In the \emph{migration} stage, the network is moved from current $\mathbf{b}$ to next $\mathbf{b}^\text{target}$ one wavelength switch at a time. A switch either removes a borrowed wavelength or activates a new borrowing; in both cases a wavelength is switched off at the leaf that loses it and on at the leaf that gains it.
A switch is performed only if it passes a \emph{transition test}. A transitional detouring matrix is computed with \textsc{2HWF} on the transitional capacities, in which the wavelength is already unavailable at the losing leaf and not yet available at the gaining one. The test succeeds if the network remains two-hop connected and any pair loads do not exceed a migration limit $\rho^{\max}_\text{th}$ (a pair already above the limit must not get worse). After a successful test: (i) the transitional matrix is applied, diverting traffic away from the wavelength to be switched off; (ii) the wavelength is switched off at the losing leaf and on at the gaining leaf, and the controller waits until it is operative; (iii) the final detouring matrix is computed by \textsc{2HWF} on the new configuration and applied. Blocked switches are retried in later passes, and switches still pending when no progress is possible are not performed (see also Appendix~VI of \cite{extended}).

\paragraph{Timescales.}
\textsc{2HWF} takes milliseconds, so it absorbs short-term traffic variation almost immediately. The migration time instead depends on the number of switches, which are executed in sequence. Each switch requires the MEMS cross-connect configuration, in the order of tens of milliseconds~\cite{lightwave2023}, plus laser activation and receiver link bring-up, which ranges from nanoseconds for laboratory burst-mode receivers~\cite{sirius2020} to seconds for standard transceivers~\cite{patronas2025}. The bound $V$ on the activations limits the migration time, at the cost of fewer optimization possibilities. The policy is therefore suited to long-term traffic variations that are much slower than the migration time. For instance, with $V=100$, a planning time of about ten seconds and a receiver synchronization time of about one second, the migration takes about two minutes, so the policy suits long-term traffic variations on the order of tens of minutes or more. 
\begin{figure}[h]
    \center
    \includegraphics[width=0.31\textwidth]{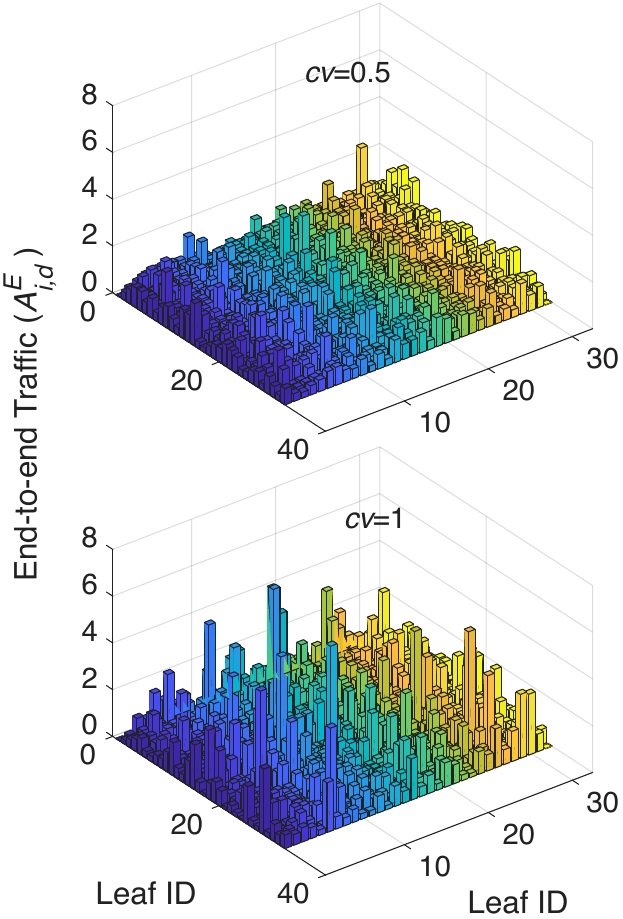}
    \caption{End-to-end traffic $\mathbf{A}^E$ for $cv=0.5$ and $cv=1$, with lognormal distribution and average 0.65.}
    \label{fig:traffic}
\end{figure}

\begin{figure*}[!t]
    \centering
    \begin{subfigure}[b]{0.31\textwidth}
        \includegraphics[width=\textwidth]{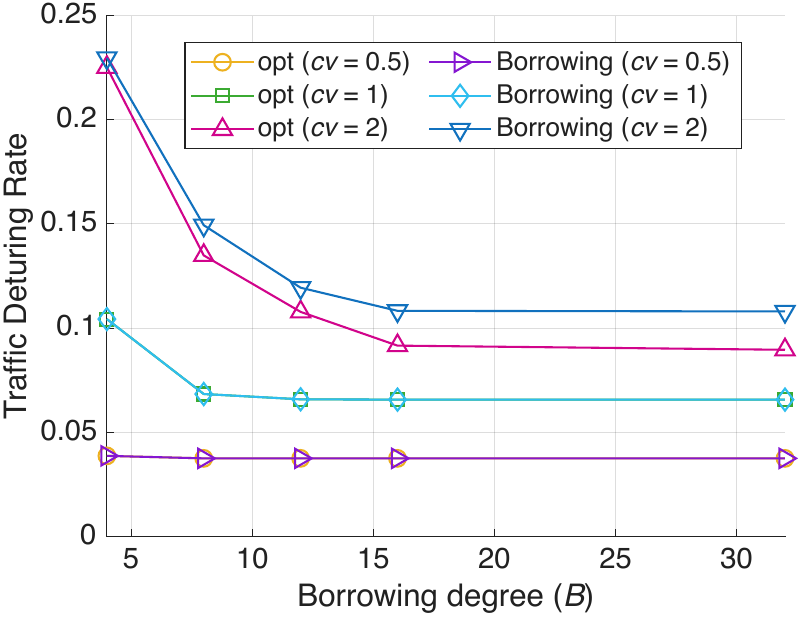}
        \caption{}
        \label{fig:milp_comp_tv}
    \end{subfigure}
    \hfill
    \begin{subfigure}[b]{0.31\textwidth}
        \includegraphics[width=\textwidth]{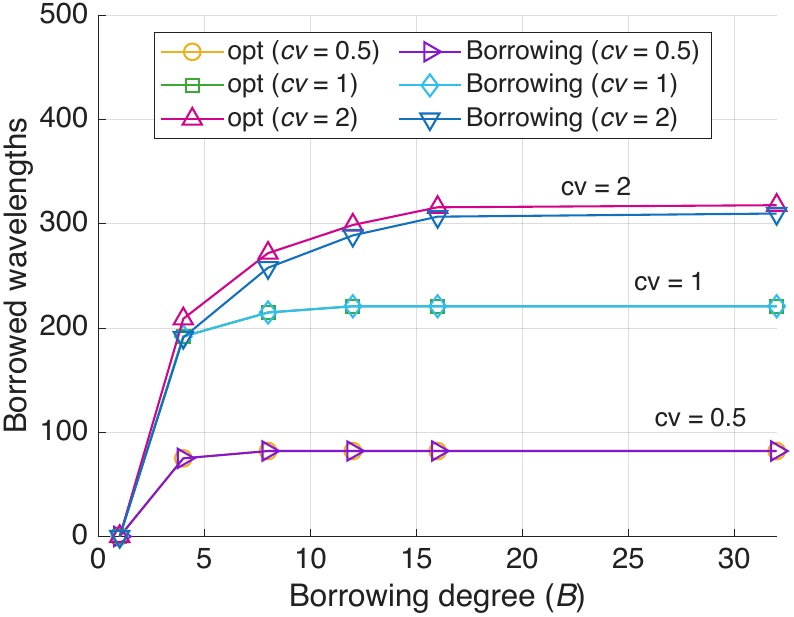}
        \caption{}
        \label{fig:milp_comp_b}
    \end{subfigure}
    \hfill
    \begin{subfigure}[b]{0.31\textwidth}
        \includegraphics[width=\textwidth]{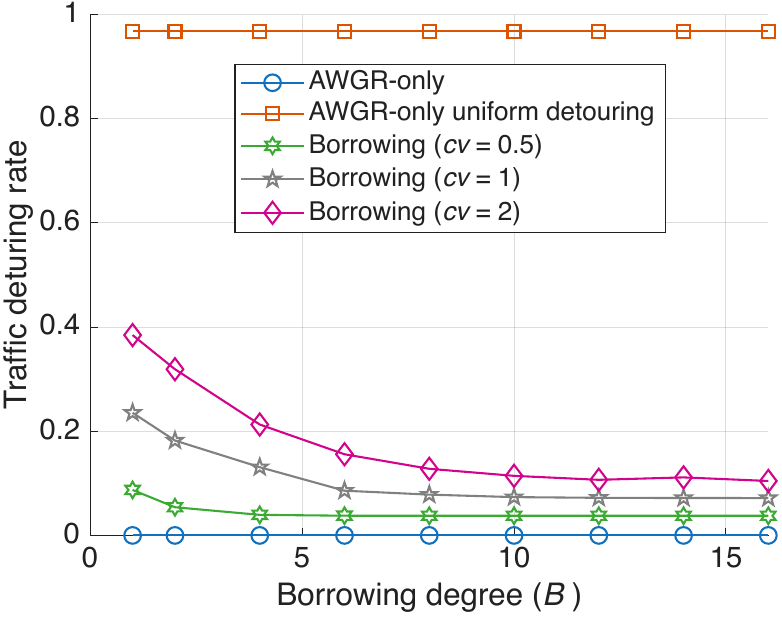}
        \caption{}
        \label{fig:detouring_rate3}
    \end{subfigure}
    \hfill
    \begin{subfigure}[b]{0.31\textwidth}
        \includegraphics[width=\textwidth]{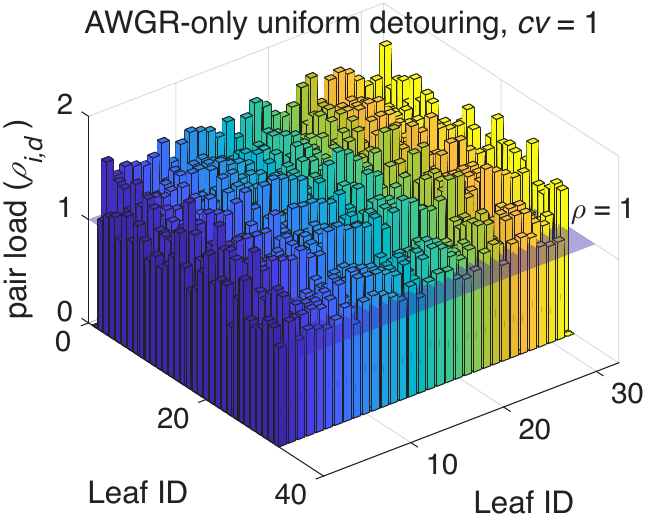}
        \caption{}
        \label{fig:rho3_awgrub}
    \end{subfigure}
    \hfill
    \begin{subfigure}[b]{0.31\textwidth}
        \includegraphics[width=\textwidth]{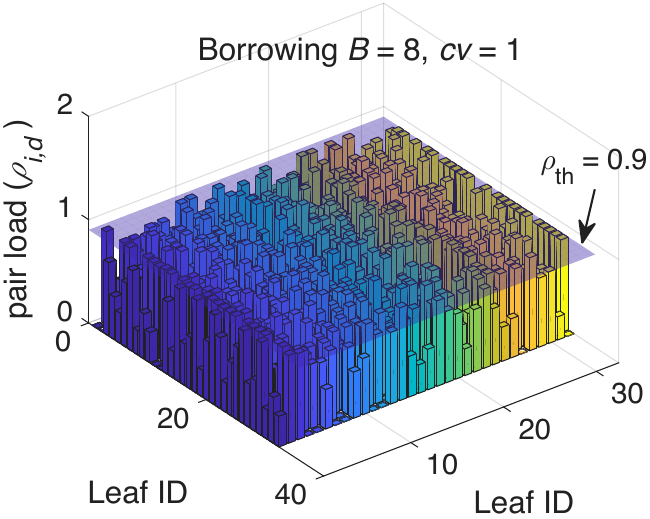}
        \caption{}
        \label{fig:rho3_borrow}
    \end{subfigure}
    \hfill
    \begin{subfigure}[b]{0.31\textwidth}
        \includegraphics[width=\textwidth]{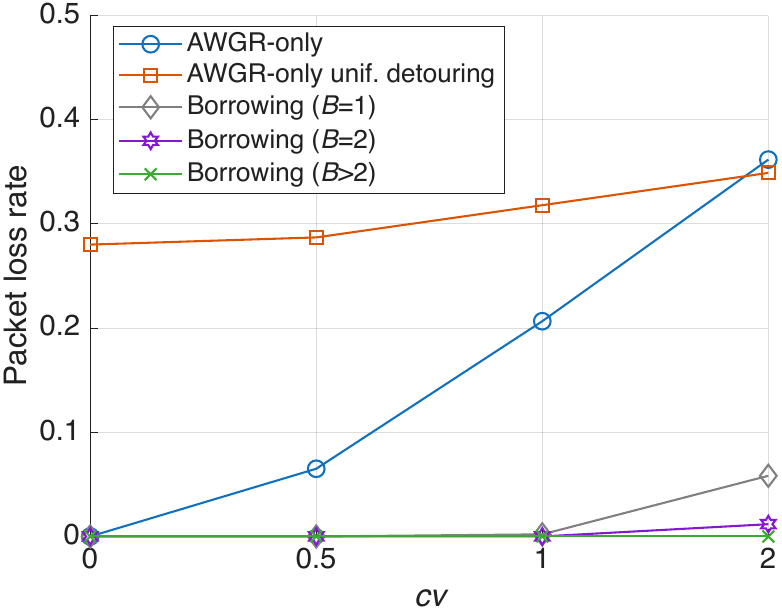}
        \caption{}
        \label{fig:packet_loss}
    \end{subfigure}
    \hfill
    \begin{subfigure}[b]{0.31\textwidth}
        \includegraphics[width=\textwidth]{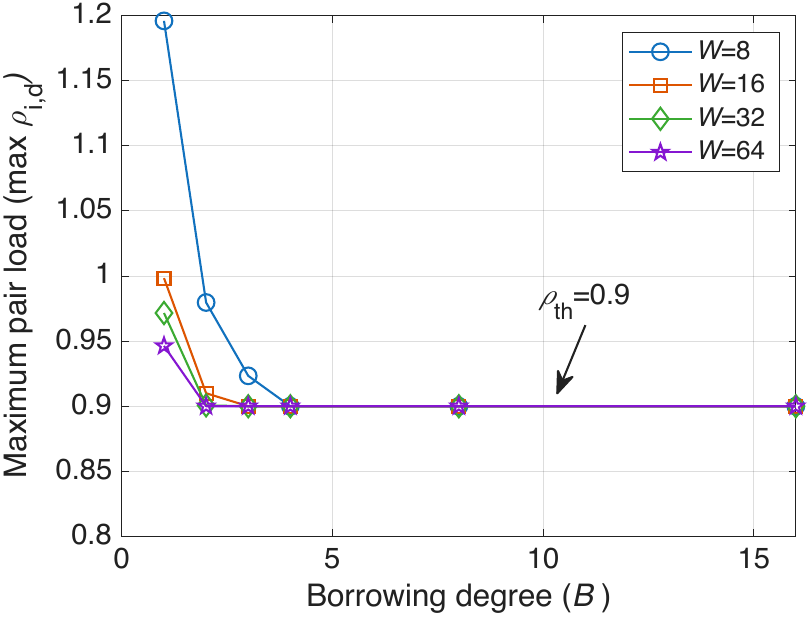}
        \caption{}
        \label{fig:rho_max_borrowing}
    \end{subfigure}
    \hfill
    \begin{subfigure}[b]{0.31\textwidth}
        \includegraphics[width=\textwidth]{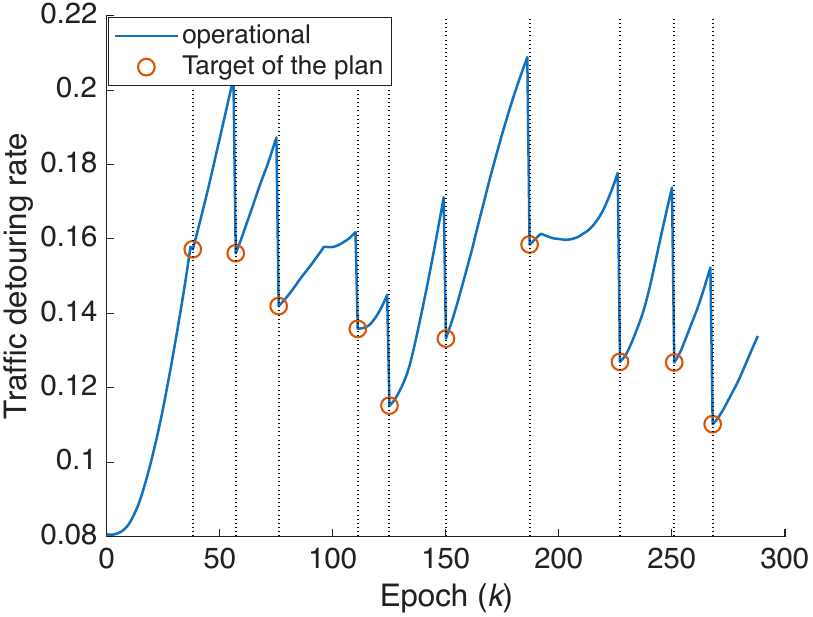}
        \caption{}
        \label{fig:dynamic_tv}
    \end{subfigure}
    \hfill
    \begin{subfigure}[b]{0.31\textwidth}
        \includegraphics[width=\textwidth]{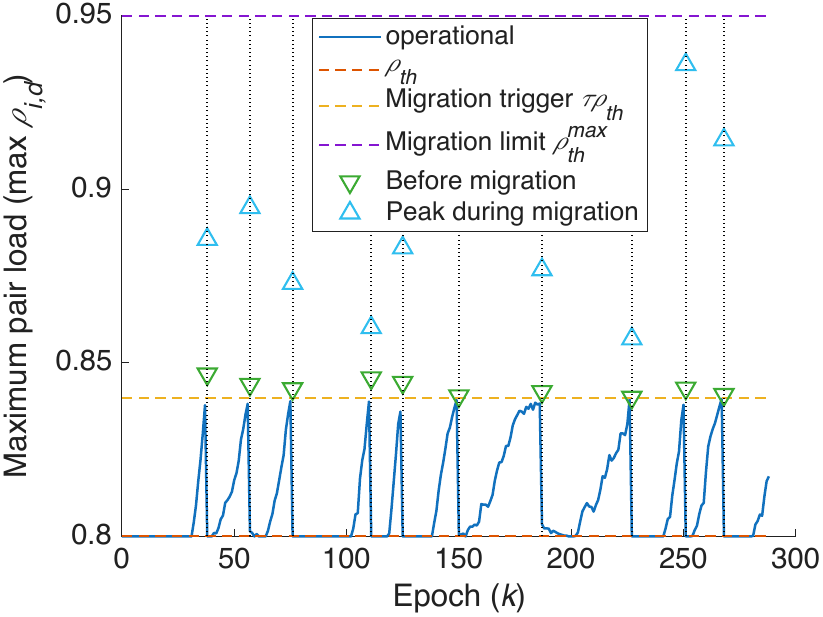}
        \caption{}
        \label{fig:dynamic_rho}
    \end{subfigure}
    \caption{Performance of the architecture with $W=32$ leaves and $\rho_\text{th}=0.9$, (a)..(f). Maximum pair load for $cv=1$ versus the number of leaves $W$ (g). Dynamic regime with $\omega=96$, $\rho_\text{th}=0.8$, $\tau=1.05$, $\rho^{\max}_\text{th}=0.95$, $V=100$ (h)-(i).}
    \label{fig:borrowing1}
\end{figure*}
\section{Performance Evaluation}
\label{sec:perf}
We developed a Python simulator implementing the heuristic in Algorithm~\ref{alg:greedy} and compared it against four baselines. The first three share the same static optical core, i.e., a cyclic AWGR with one wavelength per pair $(i,d)$, and differ only in their detouring capability; the fourth is an optimal reference:
\begin{itemize}
    \item \textit{AWGR-only} static fabric: corresponds to the architecture in \autoref{fig:arch} using only default TX modules, directly connected to the AWGR with no OxC and no detouring. It represents a static switching fabric with no traffic engineering.
    \item \textit{AWGR-only with uniform detouring} static fabric: traffic is uniformly detoured (i.e., $w_{j,i,d}=1/(W-1)\quad j\neq i$). This schedule-less approach resembles \cite{sirius2020}, though we use a dedicated laser per wavelength operating in parallel, rather than a single laser retuned at packet timescale. 
    \item \textit{AWGR-only with two-hop water-filling detouring ($B=1$)} static fabric: traffic is detoured by \textsc{2HWF}. It is completely equivalent to the borrowing architecture with $B=1$. Since this baseline shares the same optical network as the previous two, comparing them isolates the effect of the traffic-engineering solutions alone.
    \item \textit{Optimal dynamic}: borrowing $\mathbf{b}$ and detouring $\mathbf{w}$ are computed by the Gurobi optimizer solving the MILP of \autoref{sec:problem}, hence it is the best solution attainable for a given $B$ under the same objective. For $B=W$, it represents an \textit{optimal fully reconfigurable switching fabric}, in which any destination wavelength can be assigned to any source. Under the same traffic model and objective, this case upper-bounds the performance of any dynamically reconfigurable optical network with wavelength-level granularity, such as those based on Wavelength Selective Switches (WSS). It thus provides a comparison with this class of solutions that does not depend on a specific implementation.
\end{itemize}

The normalized end-to-end traffic $\mathbf{A}^E$ is modeled as a Lognormal distribution with coefficient of variation $cv$: higher $cv$ means higher traffic variability among pairs $(i,d)$ at the same average value that we considered equal to 0.65. This model simply lets us show the effectiveness of the borrowing architecture and greedy algorithm as the degree of traffic imbalance varies \footnote{Other traffic characterizations, such as the gravity model~\cite{poutievski2022jupiter}, provide the same comparative conclusions as reported in Appendix V of \cite{extended}.}. We consider three values of $cv=0.5,1,2$: traffic imbalance among pairs increases with $cv$, as shown in \autoref{fig:traffic} for $cv=0.5$ and $cv=1$. 



Figs.~\ref{fig:milp_comp_tv} and \ref{fig:milp_comp_b} compare the proposed heuristic with the optimal dynamic baseline (``opt'' in the figures) varying the borrowing degree $B$ for $W=32$ leaves (and wavelengths) with a load cap $\rho_\text{th}=0.9$. 
The heuristic and the optimal solution were run using an unoptimized Python implementation on a 2019 Intel i9 MacBook. The heuristic requires about 10\,s of computation time, versus about 600\,s for the optimal solution, which for $W=64$ and $B=8$ took more than two hours to solve versus about 90\,s of the heuristic.

For $cv=0.5$ and $cv=1$, the heuristic achieves the same performance as the optimal solution, in terms of both detouring rate (\autoref{fig:milp_comp_tv}), i.e. the detoured traffic volume $T^D$ normalized to the total end-to-end traffic $\sum A^E_{i,d}$, and number of borrowed wavelengths (\autoref{fig:milp_comp_b}). For a more marked asymmetry, as in the case of $cv=2$, the heuristic performs slightly worse. 

\autoref{fig:milp_comp_tv} shows that the detouring rate decreases as the borrowing degree increases, since greater reconfiguration flexibility can be exploited. \autoref{fig:milp_comp_b} shows that the number of borrowed wavelengths is non-decreasing in $B$ and $cv$. This confirms that both the heuristic and the optimal solution exploit a higher reconfiguration capability ($B$) to reduce traffic detouring, and that this capability is more in demand as the imbalance ($cv$) increases.

An architecture with $B=W$ is fully optically reconfigurable, since each destination wavelength can be assigned to any source. \autoref{fig:milp_comp_tv} and \autoref{fig:milp_comp_b} show that, for the lognormal traffic matrices considered ($cv\le2$), performance is already close to that of a fully reconfigurable architecture at $B=8 \ll 32$, letting the system save on the complexity of a fully reconfigurable solution. This is an empirical result that depends on the traffic. The $B$ required to approach the fully reconfigurable performance grows with the traffic imbalance $cv$ (i.e., $B=1$ is sufficient for balanced traffic ($cv=0$), while larger values are needed as $cv$ increases) and we have not evaluated extreme patterns, e.g. $cv>2$. However, measurements from production networks suggest that such patterns could be not representative of the imbalance found in practice, at least at the considered aggregation level (pods). For instance, in~\cite{poutievski2022jupiter}, the coefficient of variation of the normalized peak load $D_i$ of aggregation blocks (akin to our pods) ranges from $0.32$ to $0.56$ across ten production fabrics, and peak block-to-block traffic is well approximated by a gravity model. As detailed in Appendix V of \cite{extended}, a gravity model with this range of $D_i$ yields a coefficient of variation of the matrix entries $A^E_{i,j}$ between $0.439$ and $0.780$. Hence, among the lognormal traffic matrices considered, $cv=0.5$ and $cv=1$ bracket the values observed in production, whereas $cv\ge2$ should be regarded as a stress case, useful to probe the limits of the architecture.  

\autoref{fig:detouring_rate3} compares the detouring rate of the borrowing architecture with that of solutions based on a static AWGR fabric.  The AWGR-only configuration has no load-balancing functionality and hence no detoured traffic. 
The uniform detouring strategy used in \cite{sirius2020}, also known as Valiant load balancing \cite{valiant1982scheme}, equalizes traffic among all pairs by having every node send each packet to a randomly chosen intermediate node, which then forwards it to its final destination. This \textit{reshaping} fits the uniform topology of a static AWGR-only optical fabric, and its `schedule-less' design requires neither reconfigurable optics nor any control overhead, such as traffic-metric gathering and schedule computation. This valuable simplification has the drawback that a packet traverses the optical domain twice with probability~$(W-2)/(W-1)$, asymptotically doubling the wavelength load regardless of the traffic pattern (e.g., $cv$), which can lead to wavelength overload and packet loss. These observations are confirmed by \autoref{fig:detouring_rate3} where the detouring rate of AWGR-only with uniform load balancing is close to $30/31 \approx 97\%$, i.e., almost all traffic is detoured. 

In contrast, the borrowing architecture with \textsc{2HWF} traffic engineering is a scheduled design: it reconfigures the optical domain to minimize detouring traffic, reducing the load overhead due to double crossing of the optical core, and hence packet loss, at the cost of higher, but tunable through $B$, hardware and control complexity. Indeed, its detouring rate in \autoref{fig:detouring_rate3} is much lower than AWGR-only with uniform detouring.

Results for $B=1$ are representative of a static AWGR-only optical fabric with two-hop water-filling detouring rather than the Valiant one. Accordingly, \autoref{fig:detouring_rate3} shows that \textsc{2HWF} alone already reduces detoured traffic significantly, making it worth considering as a stand-alone traffic-engineering solution for a static, full-mesh optical fabric; further improvement requires the borrowing modules ($B>1$).

Figs.~\ref{fig:rho3_awgrub} and~\ref{fig:rho3_borrow} show the load $\rho(i,d)$ of each pair for AWGR-only with uniform detouring and borrowing architecture, respectively (the AWGR-only with no load balancing values coincide with $A^E$ in \autoref{fig:traffic} with $cv=1$). Packet loss occurs whenever $\rho(i,d)>1$, which happens for many pairs in both AWGR-only and AWGR-only with uniform detouring. Conversely, the borrowing architecture with $B=8$ keeps every pair's load below $\rho_\text{th}=0.9$, resulting in no packet loss and respecting load cap.

\autoref{fig:packet_loss} shows the packet loss rate. For $cv \leq 1$, the load-doubling drawback of uniform detouring makes its loss performance even worse than AWGR-only without load balancing. \textsc{2HWF} alone ($B=1$) avoids packet loss up to $cv=1$, but not at $cv=2$, where the traffic imbalance is severe enough to require optical reconfigurability; borrowing architecture with $B>2$ ensures zero packet loss across the whole $cv$ range tested.

\autoref{fig:rho_max_borrowing} shows an interesting scale-out behavior: as the number of leaves $W$ increases, fewer borrowing modules per leaf are needed to keep every pair's load below target and avoid loss. $B=2$ suffices for larger networks ($W=32,64$), while smaller networks need higher $B$, since a given $B$ offers more donors -- and hence more borrowing opportunities -- as the network grows. Thus, the number of borrowing modules per leaf needed to reach performance nearly equal to that of a fully reconfigurable architecture does not grow with the size of the data center. Note, however, that the number of wavelengths $W$ must still scale linearly with the number of leaves.

Figs.~\ref{fig:dynamic_tv} and \ref{fig:dynamic_rho} show the dynamic regime from a Python simulator of the dynamic control policy. As traffic load, we generated a sequence of lognormal traffic matrices $\mathbf{A}^E(k)$ with $cv=1$ and average load $0.65$. Independent matrices are drawn at the \textit{renewal epochs} $k=n\omega$, $n=0,1,2,\dots$, where $\omega$ sets the long-term traffic variability, and \textit{transitional} matrices change the traffic in between. We used $\omega=96$, $\rho_\text{th}=0.8$, $\tau=1.05$, $\rho^{\max}_\text{th}=0.95$ and activation limit $V=100$; with a control epoch of 5~min, the pod-to-pod traffic is fully renewed every 8~hours.

The detouring rate and the maximum pair load grow as the traffic drifts away from the matrix for which the optical configuration was planned, and then decrease rapidly. During the growth, the loop applies only \textsc{2HWF}, which limits the increase in detouring traffic and load caused by the progressively mismatched topology without reconfiguring the optical fabric. At specific steps, the maximum pair load obtained with \textsc{2HWF} alone would exceed the trigger $\tau\rho_\text{th}$ (\autoref{fig:dynamic_rho}); a new target configuration is then computed and the following migration phase brings the load back to the threshold $\rho_\text{th}$. A sensitivity analysis (Appendix VI of \cite{extended}) shows that a higher activation bound $V$ fits the topology to the traffic better and lengthens the interval between migrations, but increases the duration of each migration. Similarly, a longer traffic renewal period $\omega$, i.e., less dynamic traffic, lengthens the inter-migration interval.
 
A preliminary power budget of the architecture based on current technology, detailed in Appendix VII of \cite{extended}, shows that with $B=8$, the end-to-end insertion loss is approximately 27, 28, and 29~dB for $W=16,32,64$, respectively, against about 10~dB sustainable by the power budget of 400/800GBASE-DR8 SFP transceivers. DWDM-ready amplifiers in the 15--25~dB range, such as EDFAs, are therefore required after each coupler.

The cost of the architecture grows with $B$ mainly because of the transmitters. Each additional borrowing fiber requires a $W\times1$ AWG multiplexer with $W$ transmitters (e.g., SFP), but only one OxC port pair and one combiner port, so the per-leaf cost increment is $\Delta C_\text{leaf}(B)=(B-1)\,[\,W c_\text{SFP}+c_\text{MUX}+2c_\text{OxC}+c_\text{comb}\,]$, where $c_\text{SFP}$, $c_\text{MUX}$, $c_\text{OxC}$ and $c_\text{comb}$ are the costs of a transceiver, a multiplexer, an OxC port and a combiner port, respectively. The term $W c_\text{SFP}$ dominates, since optical transceivers are a substantial part of the cost of a DCN \cite{sato2018} and cost models of optical DCNs are driven by the number of transceivers \cite{sirius2020,rotos2020}. Future work could reduce the transceiver count by sharing a pool of tunable borrowing transmitters, opportunistically connected to the borrowing AWG multiplexers through a local OxC.



\section{Related Works}
\label{sec:related}
The use of all-optical switching in data centers is attracting growing interest, including from major cloud operators such as Google~\cite{lightwave2023,poutievski2022jupiter} and Microsoft~\cite{sirius2020}. Proposed architectures differ in the network tier (server-level, rack-level, block/pod-level, etc.) at which the optical fabric is deployed, which drives the traffic characteristics to be handled, the required switching speed, and the optical components. The closer to the server the optical core is placed, the more dynamic the traffic it must serve, with stringent switching/control-logic speed and component requirements; conversely, higher aggregation levels allow slower, more mature, cost-effective technologies, more complex control logic, since traffic variability is smoothed by aggregation.

Proposed solutions also differ in the maturity, and hence cost, of their optical components. Actively reconfigurable \emph{colored} fabrics---such as wavelength-selective switches (WSS) or WDM-aware OxCs with SOA gate arrays---and \emph{optical signal processing} for in-network forwarding rely on components that remain expensive and available only at limited port counts, hindering near-term deployment. By contrast, colorless MEMS-OxCs performing pure spatial switching and passive AWGRs, whose wavelength routing is fixed, are commercially available at datacenter-relevant scale, making them the more practical short-term choice.

\paragraph{Optical Core at the Server and Rack Level.}

At the finest granularity, the optical core's endpoints are directly servers or racks, whose traffic bursts must be handled with virtually no buffering. Traffic here is highly dynamic: flows are short-lived, demand changes on timescales of tens to hundreds of nanoseconds, and efficient utilization requires reconfiguration at packet or sub-packet granularity.

Sirius~\cite{sirius2020} is a prominent example. It proposes a flat, all-optical network in which a single passive layer of AWGRs replaces the entire electrical switching hierarchy above the ToR. ToR uplinks use custom tunable laser chips that encode the destination as a wavelength on a time-slot basis, enabling end-to-end reconfiguration in under 1~ns, with all-to-all connectivity via a cyclic round-robin TDMA schedule that avoids wavelength contention. To accommodate arbitrary traffic atop this uniform bandwidth topology, every packet is randomly detoured through an intermediate rack \cite{valiant1982scheme}, creating a fully balanced demand at the cost of a two-hop path, an intermediate O/E/O conversion, and near-doubling of the optical load, since almost all packets traverse wavelengths twice. The AWGR core is passive, with no active reconfiguration or optical signal processing -- routing follows purely from the transmitter's wavelength choice -- but the architecture requires custom photonic integrated circuits for the nanosecond tunable laser, and its lack of reconfigurability forces uniform two-hop detouring regardless of actual traffic skew. 

OPSquare~\cite{miao2016} and its multi-level extension HFOS~\cite{hfos} push optical switching to the ToR level using fast \emph{colored} (WDM-aware) optical packet switches with nanosecond-scale reconfiguration. Each switch combines AWGs with SOA-based $1{\times}F$ broadcast-and-select gate arrays, forwarding packets within a group of $F$ ToRs by extracting an in-band RF-tone \emph{optical label} to control the SOA gates -- genuine \emph{optical signal processing}. Since no prior scheduling is performed, contention causes ALOHA-like packet loss, recovered via ACK/NACK retransmission from electrical buffers. HFOS scales this to multiple parallel switch levels, reaching tens of thousands of servers under the same colored, processing-based paradigm. Both remain technologically demanding -- label processors are research-grade and available only at small port counts -- and suffer non-negligible loss and low throughput under load due to lack of contention avoidance.

ROTOS~\cite{rotos2020} extends OPSquare~\cite{miao2016} with a reconfigurable ToR switch that dynamically reallocates WDM transceivers and a colored WSS between intra- and inter-cluster traffic under SDN control, steering wavelengths via WSS according to the observed traffic ratio. It remains a packet-level solution using the same colored, SOA-based, label-processing switches as OPSquare, and additionally requires a per-ToR WSS, adding cost and a scalability constraint proportional to the node count $N$.

PULSE~\cite{pulse} builds a packet-level all-optical network around $x^2$ passive $N{\times}N$ star couplers ($N$ servers/rack, $x$ racks). Each server has $x$ transceivers, one per rack, each with $B$ banks of continuously-on tunable DS-DBR lasers covering $W{=}N$ wavelengths~\cite{ward2005widely}; SOA-gated laser outputs open only during the reserved time slot, and a star coupler broadcasts to the destination rack. Wavelength/slot assignment is precomputed by per-rack schedulers to avoid contention. Its scalability is limited by star-coupler splitting loss, which grows as $10\log_{10}(N)$\,dB with rack size and requires SOA amplification at every transceiver, plus $x$ nanosecond-scale tunable transceivers per server -- a costly, complex requirement for off-the-shelf hardware.

\paragraph{Optical Switching at the Pod Level.}

A complementary class of architectures places the optical fabric at a higher level, interconnecting aggregation points such as leaf nodes or Pods. Here the smoother, slowly varying traffic enables slower but more mature and cost-effective technologies: colorless MEMS-based OxC, colored WSS, and passive AWGR.

Google's Jupiter~\cite{poutievski2022jupiter} is a prominent industrial deployment replacing the electrical spine layer with a datacenter network interconnection layer built on colorless MEMS-based OxCs that connect aggregation blocks via pure spatial switching, with no wavelength awareness or optical signal processing. Traffic engineering reroutes traffic on timescales of seconds to minutes, whereas the OxC topology is reconfigured rarely, on the scale of weeks. Lightwave Fabrics~\cite{lightwave2023} describes the same OxC approach in production, together with custom bidirectional transceivers, for datacenter networks and for machine-learning superpods.

FleX-LION~\cite{liu2020architecture}~\cite{xiao2019flex}, which inspired our reconfigurability/complexity trade-off, is a silicon-photonic switch fabric combining an AWGR, $B$ MRR add-drop filters per port, and an $N{\times}N$ MRR crossbar. Up to $B$ wavelengths of a leaf can be dropped from their default AWGR path and \emph{rerouted} to another output, at the expense of the default connectivity. Our architecture instead \emph{adds} wavelengths by borrowing idle slots of other leaves, leaving the default AWGR connectivity untouched, and uses only off-the-shelf passive AWGR, combiners and a colorless MEMS-OxC, with no wavelength-selective element. The authors of~\cite{hyperflex} extend FleX-LION to Hyper-FleX-LION, which operates at rack level with an aggregation-like paradigm: a passive AWGR is combined with colored $1{\times}N$ WSSs at each rack's TX/RX, steered by the SDN controller on a millisecond timescale. This requires $2N^2$ WSSs for $N$ racks, whose cost and commercial port count ($N \approx 20$) bound its scalability.

The large-scale fast optical circuit switch of~\cite{sato2018} demonstrates colorless MEMS-based OxC at hundreds of ports with acceptable loss and millisecond switching, confirming the viability of this technology class; a multi-stage Thin-CLOS AWGR fabric~\cite{proietti2018thinclos} offers an alternative, passive route to the same scale. Reconfiguring an OxC-based fabric over time introduces a connection defragmentation problem, studied in~\cite{dong2025risk}. Ring-based aggregation fabrics from grouped ROADMs with shared amplification have similarly been proposed~\cite{zhao2017roadmring}, and~\cite{slotted} explores sub-wavelength TDMA resource allocation at the optical layer, a technique also optionally supported by the proposed architecture.

\paragraph{The Proposed Architecture.}

The wavelength-borrowing architecture targets the same pod-level design space as Jupiter, Hyper-FleX-LION, and ROTOS, but pursues a distinctive \textbf{complexity-reconfigurability trade-off grounded in technological maturity}. Its central optical fabric relies exclusively on a passive AWGR and a colorless MEMS-based OxC performing pure spatial switching --- both well-established, commercially available technologies --- with no actively reconfigurable colored components and no optical signal processing. This contrasts with OPSquare and HFOS, which require colored WDM-aware OxCs with SOA gate arrays and optical label processors, and with Hyper-FleX-LION and ROTOS, which rely on per-node WSSs. In place of the WSS, the proposed architecture uses a simple passive combiner. Unlike ROTOS and Hyper-FleX-LION, moreover, activating or releasing a borrowed wavelength never requires reconfiguring the destination leaf, which keeps receiving on the same AWGR output port throughout, apart from the clock recovery of its receiver on the new source signal.

Resource allocation flexibility is instead controlled through the single borrowing degree $B$, providing a tunable trade-off between hardware complexity and reconfiguration flexibility: \autoref{sec:perf} shows that performance nearly equal to that of a fully reconfigurable architecture is already achieved at a small $B$ regardless of network scale, and that the number of borrowing lines required per leaf can even decrease as the data center grows, in contrast with architectures such as Hyper-FleX-LION whose per-node WSS count scales with the network size.

Since wavelengths are reconfigured one at a time by the MEMS-based OxC, the architecture suits pod-to-pod demands that vary over tens of minutes or more, consistent with aggregation-level deployment~\cite{poutievski2022jupiter}. Finally, the paper couples the architecture with a MILP formulation of the joint wavelength-assignment and detouring problem and a greedy heuristic, built around the \textsc{2HWF} subroutine, that also works as a stand-alone detouring solution for other optical fabrics.
Appendix III of \cite{extended} reports a comparative table of the considered architectures.

\section{Conclusions}
\label{sec:conclusions}
This paper presented a wavelength-borrowing architecture for spine-leaf optical data center networks, in which idle or lightly loaded wavelengths at one leaf are dynamically reallocated to a leaf with higher demand. The borrowing degree $B$ exposes the complexity/reconfigurability tradeoff as a single tunable parameter, from a static core ($B=1$) to a fully reconfigurable fabric ($B=W$). Combined with the proposed \textsc{2HWF} traffic-engineering heuristic, results show that a moderate, scale-independent borrowing degree, e.g., $B=8$, achieves performance nearly equal to that of a fully reconfigurable fabric, and that fewer borrowing lines per leaf are needed as the network scales out.
A dynamic control loop absorbs gradual traffic variations through traffic engineering alone and reconfigures the optical layer only when pair loads are about to exceed the tolerated threshold, reconfiguring one wavelength at a time without traffic interruption. Its sequential migration makes the architecture suited to pod-to-pod demands that vary on timescales of tens of minutes or more. Future work can parallelize the migration to relax this timescale, and exploit the transparency to time division multiple access (TDMA) to refine the borrowing unit below a whole wavelength without changing the optical fabric.

\bibliography{sample}

\clearpage
\appendix
\section*{APPENDIX I: Architectural Extension}
\label{sec:ext}
\subsection{Small Data Center}
The foregoing description assumed that the number of leaf nodes $L$ equals the number of wavelengths $W$. If the required number of leaves $L$ is smaller than $W$, the spine optical fabric remains unchanged, and the default wavelengths of the $W-L$ absent leaves can simply be borrowed by the existing leaves.

\subsection{Large Data Center}
\begin{figure}[t]
      \centering
      \includegraphics[scale=0.55]{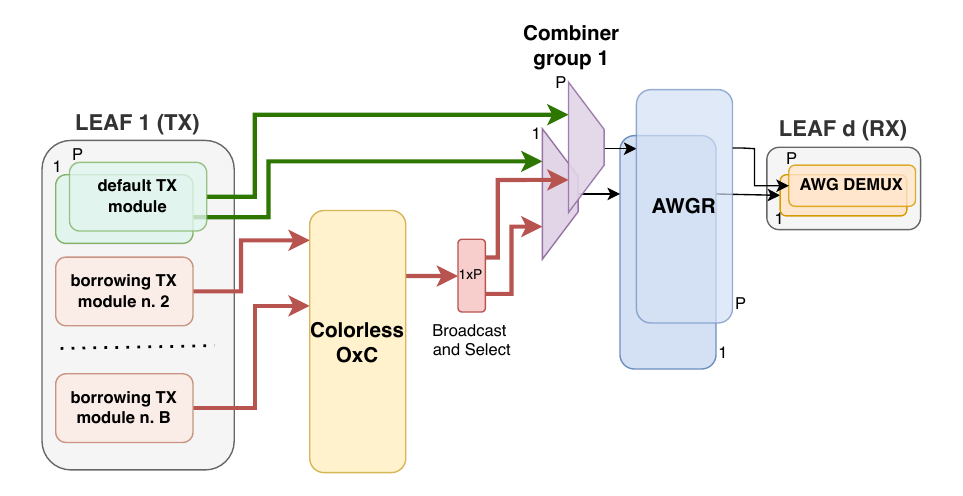}
      \caption{Capacity scaling with $P$ parallel AWGR-routed layers}
      \label{fig:scaling}
\end{figure}
Hyperscale data centers can require very large bisection bandwidth and large numbers of nodes to interconnect. In the proposed architecture, the maximum number of leaves is $W$, and the bidirectional bisection bandwidth in the balanced configuration is $W^2 S / 2$, where $S$ is the per-wavelength bitrate. For instance, with $W=64$ and $S=400$~Gbit/s, the resulting bisection bandwidth is about 0.8~Pbit/s \cite{sato2018,lightwave2023,ieee8023dj2023}.

\paragraph{Bandwidth scaling}
If this bisection bandwidth is insufficient but the number of leaf nodes $W$ is adequate, the architecture can be \emph{layered} as shown in \autoref{fig:scaling}, where only the components related to transmitting leaf~1 and receiving leaf~$d$ are depicted. Specifically, the architecture provides $P$ parallel AWGR-routed layers carrying both default and borrowed wavelengths. The colorless OxC is shared across all $P$ layers, and its size does not grow with $P$, thereby removing potential scale-out limitations imposed by the unavailability of large OxC switches. The resulting bisection bandwidth scales by a factor of $P$.

In this scaling scheme, each leaf~$i$ has the usual $B-1$ borrowing fibers connected to the central OxC, but $P$ parallel default TX modules, resulting in $P$ output default fibers each carrying $W$ default wavelengths toward the $W$ remote leaves. The $P$ default fibers are connected to group~$i$ of $P$ parallel combiners, whose output fibers are in turn connected to $P$ parallel AWGRs. Output ports $d$ of these AWGRs are connected to $P$ parallel AWG demultiplexers and WDM receivers at destination leaf~$d$.
The colorless OxC remains unchanged and routes the $B-1$ borrowing fibers per node to the combiner groups. Routing within each group to a specific combiner is managed by a $1{\times}P$ optical switch, e.g., implemented with broadcast-and-select technology~\cite{rotos2020}, configured by the SDN controller. In this way, a borrowing fiber can opportunistically access resources from any AWGR layer.

\paragraph{Node scaling}
\begin{figure}[t]
      \centering
      \includegraphics[scale=0.52]{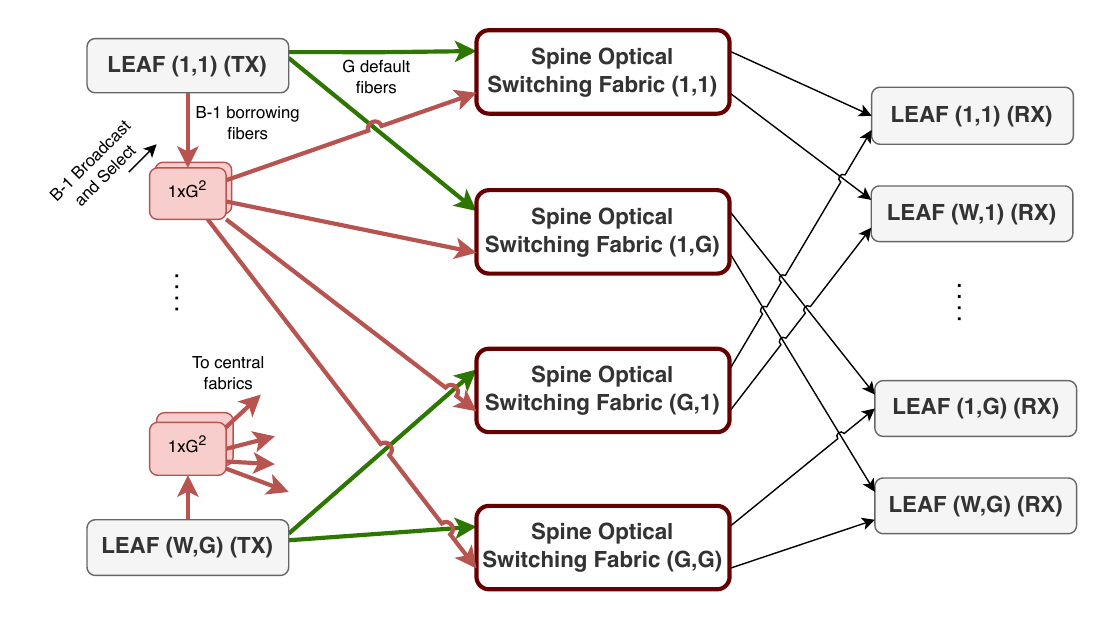}
      \caption{Node scaling with cross-connecting optical fabrics. Leaf~$(i,j)$ denotes leaf~$i$ of group~$j$; central fabric~$(t,r)$ connects TX leaves of group~$t$ to RX leaves of group~$r$.}
      \label{fig:scaling2}
\end{figure}
When more than $W$ leaf nodes are required, the architecture can be extended as shown in \autoref{fig:scaling2}.
The nodes are partitioned into $G$ groups of at most $W$ nodes each. Communication within the same group and between different groups is handled by $G^2$ distinct spine optical switching fabrics, one per ordered source--destination group pair. For example, central fabric~$(1,1)$ interconnects TX leaves of group~1 with RX leaves of group~1, while central fabric~$(1,G)$ cross-connects TX leaves of group~1 with RX leaves of group~$G$.

\autoref{fig:scaling2} depicts a representative subset of leaves and interconnections: transmitting leaf~1 of group~1, identified as leaf~$(1,1)$ (TX); central fabric~$(1,G)$ connecting group~1 to group~$G$; and receiving leaf~1 of group~$G$, identified as leaf~$(1,G)$ (RX).

Each TX leaf has $G$ distinct default fibers, each sourced by a dedicated default TX module. The $g$-th default fiber carries $W$ wavelengths destined for the RX leaves of group~$g$ and is connected to the spine fabric serving that source--destination group pair. In addition, each TX leaf outputs $B-1$ borrowing fibers, which carry wavelengths lent by donor leaves. Since a borrowed wavelength may interconnect leaves belonging to any pair of groups, each borrowing fiber must be steerable to the appropriate spine fabric. This is accomplished by a dedicated $1{\times}G^2$ broadcast-and-select switch per borrowing fiber, configured by the SDN controller%
\footnote{The fan-out of these switches can be reduced by restricting the set of central fabrics from which each borrowing fiber may borrow resources.}.
Finally, bisection bandwidth can be further increased by combining both scaling approaches.

\section*{APPENDIX II: Derivation of the Two-Hop Absorbed Traffic}
\label{sec:app-Fs}
This appendix motivates \eqref{eq:Fs}, the amount of traffic $F_{\theta,j,i,d}$ that a two-hop path $(j,i,d)$ absorbs at water level $\theta$.

Let $\Delta_{j,i}=\rho^{new}_{j,i}-\rho^{old}_{j,i}$ and $\Delta_{i,d}=\rho^{new}_{i,d}-\rho^{old}_{i,d}$ denote the load increments induced on the two hops by the detoured traffic. Since the same physical traffic $F_{\theta,j,i,d}$ traverses both hops of the path,
\begin{equation}
    F_{\theta,j,i,d}=\Delta_{j,i}\,c_{j,i}=\Delta_{i,d}\,c_{i,d}.
    \label{eq:app-same-flow}
\end{equation}

For a single link taken in isolation, the largest amount of traffic it could absorb without exceeding the target level $\theta$ is obtained by setting $\rho^{new}=\theta$, i.e.,
\begin{equation}
    F^{j,i}_{\max}=(\theta-\rho^{old}_{j,i})\,c_{j,i}, \qquad
    F^{i,d}_{\max}=(\theta-\rho^{old}_{i,d})\,c_{i,d}.
    \label{eq:app-fmax}
\end{equation}

Because $F_{\theta,j,i,d}$ in \eqref{eq:app-same-flow} is the same quantity on both hops -- not the same $\rho$ increment -- it cannot exceed either per-link limit in \eqref{eq:app-fmax}. The more constrained of the two hops therefore saturates first and bounds the absorbed traffic:
\begin{equation}
    F_{\theta,j,i,d}=\min\big(F^{j,i}_{\max},\,F^{i,d}_{\max}\big).
    \label{eq:app-min}
\end{equation}
The hop attaining the minimum in \eqref{eq:app-min} reaches $\rho^{new}=\theta$ exactly, while the other hop remains below $\theta$, since it is bounded away from saturation by construction. Hence $\max(\rho^{new}_{j,i},\rho^{new}_{i,d})=\theta$, consistently with the two-hop load definition $\rho_{j,i,d}=\max(\rho_{j,i},\rho_{i,d})$.

Finally, if a hop is already loaded past $\theta$ (i.e., $\rho^{old}>\theta$), its term in \eqref{eq:app-fmax} is negative and must be clamped to zero, since a link cannot absorb a negative amount of traffic; this is the case of path~3 in \autoref{fig:wf}, which is already more loaded than $\theta$ and therefore cannot be filled. Applying this clamp to \eqref{eq:app-min} yields \eqref{eq:Fs}.

\section*{APPENDIX III: Comparison of Related Architectures}
\label{sec:app-comparison}
\autoref{tab:related2} shows a comparison between the proposed architecture and the all-optical data center architectures discussed in \autoref{sec:related}, in terms of network tier, switching, transmission and reception technologies, optical multiplexing, resource allocation flexibility, optical processing, and technology maturity and cost.

\begin{sidewaystable*}[p]
\centering
\caption{Comparison of all-optical datacenter network architectures.
``Optical processing'' indicates in-network forwarding decisions made
in the optical domain (e.g., optical label extraction, SOA gating).
``Rack group'' denotes a cluster of racks served by a common
aggregation point (leaf node, aggregation block, or cluster switch).
\textbf{Opt.\ tech.\ maturity \& cost} rates the key switching
technology as \emph{commercial} (off-the-shelf, volume pricing) or
\emph{laboratory} (custom/prototype, limited availability).}
\label{tab:related2}
\begin{tabular}{|>{\bfseries}p{2.4cm}|p{3.0cm}|p{3.0cm}|p{3.0cm}|p{3.0cm}|p{3.0cm}|p{3.0cm}|}
\hline
\textbf{Property}
& Sirius~\cite{sirius2020}
& PULSE~\cite{pulse}
& OPSquare~\cite{miao2016}, HFOS~\cite{hfos}, ROTOS~\cite{rotos2020}
& Hyper-FleX-LION~\cite{hyperflex}
& Jupiter~\cite{poutievski2022jupiter}, Lightwave Fabrics~\cite{lightwave2023}
& This work\\
\hline

Network tier
& Rack/Server-level
& Server-level
& Rack-level
& Rack-level
& Rack-group-level
& Rack-group-level\\
\hline

Key switching tech.
& Passive AWGRs
& Passive star coupler
& AWG + SOA gates + WSS (ROTOS only)
& Passive AWGR + colored WSS
& Colorless MEMS OXC
& Colorless MEMS OXC + passive AWGR + Combiners\\
\hline

TX technology
& Fast tunable WDM ($<$1\,ns)
& Fast tunable WDM + SOA/AWG bank
& Fixed-wavelength WDM
& Fixed-wavelength WDM
& Not mandatory (whole fibers switched)
& Fixed-wavelength WDM\\
\hline

RX technology
& Burst-mode fixed-$\lambda$ WDM + phase-caching
& Burst-mode fast tunable WDM + SOA/AWG bank or coherent RX
& Burst-mode fixed-$\lambda$ WDM
& Fixed-$\lambda$ WDM
& Not mandatory (whole fibers switched)
& Fixed-$\lambda$ WDM\\
\hline

Optical Multiplexing
& WDM+TDMA
& WDM+TDMA
& WDM
& WDM
& Not mandatory (whole fibers switched)
& WDM \\
\hline

Resource allocation flexibility
& Low---Static cyclic WDM/TDMA schedule.
& High---Per-packet $\lambda$+slot alloc.
& High/Medium---ALOHA-like on predefined $\lambda$ (OPSquare/HFOS); SDN/WSS-driven $\lambda$ reallocation (ROTOS).
& Medium---$\lambda$ reconfiguration.
& Low---fiber-level reconfiguration
& Tunable, static-to-high---$\lambda$ reconfiguration via the borrowing degree $B$, from a static core ($B=1$) to full reconfigurability ($B=W$)\\
\hline

Optical processing
& No
& No
& Yes
& No
& No
& No\\
\hline

Opt. tech. maturity \& cost (key switch elem.)
& Passive AWGR: \textbf{commercial}, low cost. Fast tunable laser: \textbf{laboratory}, high cost wrt fixed laser.
& Star coupler: \textbf{commercial}, low cost. Fast DS-DBR + SOA bank: \textbf{laboratory}/early comm., medium-high cost.
& AWG + SOA gates: \textbf{commercial}, std.\ cost. Label processor: \textbf{laboratory}, small port count. WSS (ROTOS): \textbf{commercial}, med.-high. Overall: laboratory-grade, medium-high cost.
& Passive AWGR: \textbf{commercial}, low cost. Colored WSS: \textbf{commercial}, med.-high. Overall: medium cost.
& 3D MEMS OXC: \textbf{commercial}, med.\ cost ($320{\times}320$, ms switching). WDM TRX (CWDM4): \textbf{commercial}, low cost.
& MEMS OXC: \textbf{commercial}, med.\ cost. Passive AWGR + combiners: \textbf{commercial}, low cost. Fixed-$\lambda$ WDM TRX: \textbf{commercial}, low cost.\\
\hline
\end{tabular}
\end{sidewaystable*}

\section*{APPENDIX IV: Complexity Analysis of the Algorithms}
\label{sec:alg-complexity}
This appendix evaluates the computational complexity of Algorithms~\ref{alg:wf} and~\ref{alg:greedy} as a function of the number of leaves $W$.

\paragraph{Complexity of 2HWF (Algorithm~\ref{alg:wf})}
Lines 1--6 compute the capacities $c_{i,d}$ from $\mathbf{b}$, the traffic terms $A^D_{j,d}$ and $A^{H0}_{j,d}$, and the initial loads, and sort the set $S$. Their cost is $O(W^3)$, dominated by the computation of $\mathbf{c}$, plus $O(W^2\log W)$ for the sorting.
The complexity of the algorithm is dominated by the loop at line~7, which iterates over at most $W(W-1)=O(W^2)$ leaf pairs. For each pair, the set $R_{j,d}$ is built in $O(W)$. The water level $\theta$ is computed once per pair by solving \eqref{eq:L}: since $F_{\theta,j,d}$ in \eqref{eq:F} is piecewise linear in $\theta$ with at most $2\lvert R_{j,d}\rvert\le 2W$ breakpoints, the solution is found by sorting the breakpoints, with complexity $O(W\log W)$, and by a single linear sweep over them. The inner loop at line~11 over $i\in R_{j,d}$ performs $O(1)$ operations for each intermediate leaf, i.e., $O(W)$ in total. Each iteration of the outer loop thus costs $O(W\log W)$, and the loop costs $O(W^2)\cdot O(W\log W)=O(W^3\log W)$. Finally, computing $T^D$ costs $O(W^3)$. The overall complexity of 2HWF is therefore
\begin{equation}
  T_{\textsc{2HWF}}(W)=O(W^3\log W)
\end{equation}

\paragraph{Complexity of Algorithm~\ref{alg:greedy}}
Phase~1 executes 2HWF once, hence its cost is $T_{\rm P1}=O(W^3\log W)$.

In Phase~2, a new round of the greedy loop starts each time a borrowing is applied, so the number of rounds is at most $N_{\rm app}+1$, where $N_{\rm app}$ is the number of applied borrowings. We bound $N_{\rm app}$ using the hardware constraints. By~\eqref{eq:optc2}, each of the $W$ borrowers can borrow from at most $B-1$ donors and, from each donor, at most one wavelength per destination, i.e., at most $W-1$ wavelengths. Thus $N_{\rm app}\le W(B-1)(W-1)$. Furthermore, a donor must retain at least one default wavelength to guarantee the two-hop connectivity in~\eqref{eq:optc5}, so it can lend at most $W-1$ of its $W$ default wavelengths, and by~\eqref{eq:optc3} each wavelength is lent at most once, which gives $N_{\rm app}\le W(W-1)$. Therefore
\begin{equation}
    N_{\rm app}\le\min\{W(W-1)(B-1),\ W(W-1)\}.
\end{equation}
In each round, building and sorting the set of borrowing candidates $\mathcal{C}$ costs $O(\lvert\mathcal C\rvert\log\lvert\mathcal C\rvert)=O(W^3\log W)$. Each candidate, taken in sorted order, is first checked against~\eqref{eq:optc1}--\eqref{eq:optc5}, at a cost that does not exceed that of 2HWF. If the check is passed, 2HWF is executed to evaluate the candidate. Let $E$ denote the total number of these tentative 2HWF evaluations. Each candidate is either applied or inserted in the skip set $K$, and a triple in $K$ is never considered again; thus $E\le N_{\rm app}+\lvert K\rvert=O(W^3)$, independently of the number of rounds and the evaluation complexity is $O\big(E\,W^3\log W\big)$. Moreover, every applied borrowing results from one evaluation, so $N_{\rm app}\le E$ and the number of rounds is at most $E+1$. The cost of building and sorting the candidates over all rounds, at most $(E+1)\,O(W^3\log W)$, is therefore of the same order as that of the evaluations. The resulting complexity $T_{\rm P2}$ of Phase~2 is
\begin{equation}
    \label{eq:complexity}
    T_{\rm P2}=O\big(E\,T_{\textsc{2HWF}}\big)=O\big(E\,W^3\log W\big)\le O\big(W^6\log W\big).
\end{equation}

Phase~3 considers at most $W$ leaves $j$. For each of them, it tests at most $W-1$ candidate borrowers $i\neq j$ against~\eqref{eq:optc1}--\eqref{eq:optc5}, with a cost $O(W)$ per test when~\eqref{eq:optc5} is verified incrementally, and executes 2HWF once. Its complexity is therefore $T_{\rm P3}=O\big(W\,(W^2+T_{\textsc{2HWF}})\big)=O(W^4\log W)$.

Since $T_{\rm P1}$ and $T_{\rm P3}$ are dominated by $T_{\rm P2}$ in the worst case, the complexity of Algorithm~\ref{alg:greedy} is given by~\eqref{eq:complexity}. 

\begin{figure}[h]
      \centering
      \includegraphics[scale=0.55]{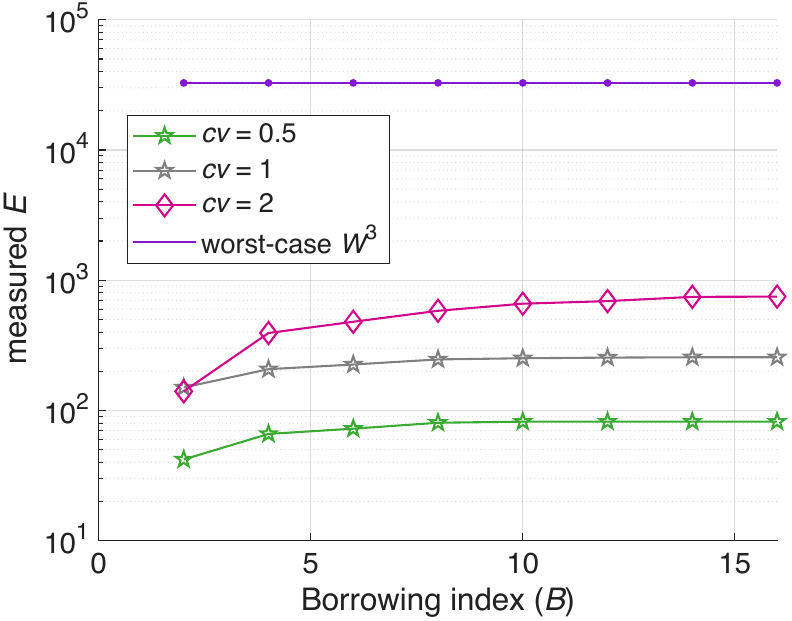}
      \caption{Measured number of 2HWF evaluations ($E$) for W=32, $\rho_\text{th}=0.9$ versus $B$}
      \label{fig:n_wf}
\end{figure}

\begin{figure}[h]
      \centering
      \includegraphics[scale=0.55]{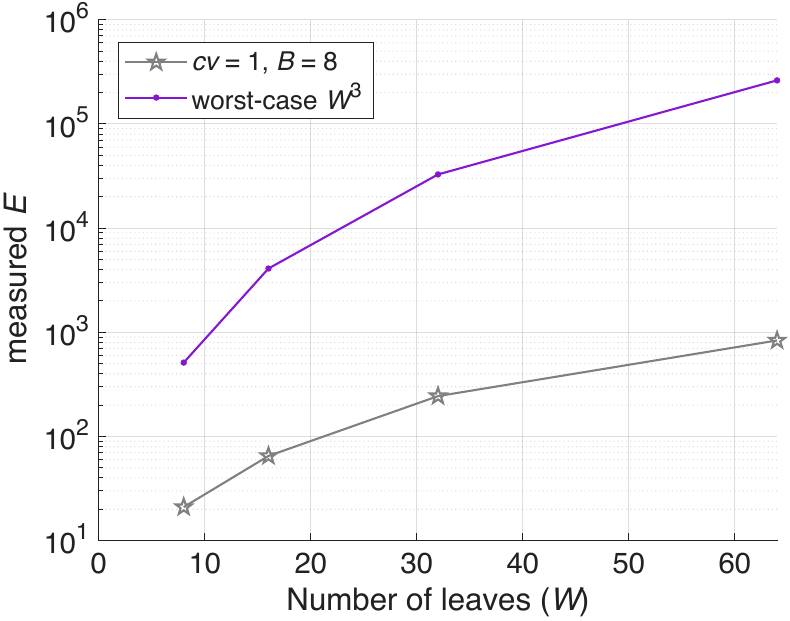}
      \caption{Measured number of 2HWF evaluations ($E$) for W=32, $cv=1$, $\rho_\text{th}=0.9$ versus $W$}
      \label{fig:n_wf_n}
\end{figure}

In practice, $E$ is much smaller than $W^3$, since only triples whose pair $(i,d)$ has traffic to detour ($A^D_{i,d}>0$) are candidates. For instance, \autoref{fig:n_wf} shows the measured $E$ for $W=32$ as a function of the borrowing degree $B$ and of log-normal $cv$, compared with the worst-case value $W^3=32768$. $E$ increases with $B$, since more borrowing options can be attempted, but it saturates for $B\gtrsim 8$ at about $80$, $260$ and $750$ for $cv=0.5$, $1$ and $2$, respectively, i.e., more than on order of magnitude below $W^3$. $E$ also increases with $cv$, since a greater traffic imbalance requires more attempts to rearrange the optical topology. \autoref{fig:n_wf_n} confirms that $E$ remains well below the $W^3$ bound as the data center size increases, with a slightly widening gap.   Hence, in practice the cost of Phase 2 is $O(E\,W^3\log W)$ with $E\ll W^3$. 


\section*{APPENDIX V: Gravity Model}
\label{sec:gravity}

This appendix reports the evaluation with gravity-model traffic matrices $A^E_{i,j}$, which complements the lognormal pair-demand model used in Section~\ref{sec:perf}. The gravity model was validated on production traces in~\cite{poutievski2022jupiter}.

Let $D_i$ be the aggregate normalized traffic demand of leaf $i$, assumed symmetric in the input and output directions and including the intra-leaf traffic. According to the gravity model, the normalized traffic demand from leaf $i$ to leaf $j$ is
\begin{equation}
  A^E_{i,j} = \frac{D_i\,D_j}{\sum_{k=1}^{W} D_k}, \qquad i,j=1,\dots,W,
  \label{eq:gravity}
\end{equation}
The term $A^E_{i,i}=D_i^2/\sum_k D_k$ is the intra-leaf traffic, which does not cross the optical spine. The traffic offered by leaf $i$ to the spine is therefore
\begin{equation}
  \sum_{j\neq i}A^E_{i,j}=D_i\left(1-\frac{D_i}{\sum_{k}D_k}\right),
\end{equation}
and only the entries $A^E_{i,j}$ with $i\neq j$ are used in the evaluation.

We assume that the $D_i$ are i.i.d.\ lognormal random variables with mean $m_b$ and coefficient of variation $cv_b$, i.e.,
\begin{equation}
  D_i \sim \mathrm{LogN}(\mu,\sigma^2), \qquad
  \sigma^2=\ln\!\left(1+cv_b^2\right), \quad
  \mu=\ln m_b-\frac{\sigma^2}{2}.
\end{equation}
The lognormal distribution is positive and right-skewed with a heavy tail, which is consistent with the observation in~\cite{poutievski2022jupiter} that a small number of blocks (leaves, in our terminology) contributes most of the offered load.

Reference~\cite{poutievski2022jupiter} reports that, across ten production fabrics, the coefficient of variation of the 99th-percentile (peak) offered load of the blocks, normalized to their capacity, ranges from $0.32$ to $0.56$. It reports neither the distribution of the demands nor the coefficient of variation of their average values. In the absence of this information, we interpret $D_i$ as the demand measured during the peak period and set $cv_b\in\{0.32,\,0.56\}$, the two extremes of the reported range. The scale of $D_i$ is chosen so that the average of the off-diagonal entries of $A^E$ is $0.65$, and matrices whose inbound load at some leaf exceeds the leaf capacity are discarded and redrawn. 

\begin{figure*}[h]
    \centering
    \begin{subfigure}[b]{0.33\textwidth}
        \centering
        \includegraphics[height=4.4cm]{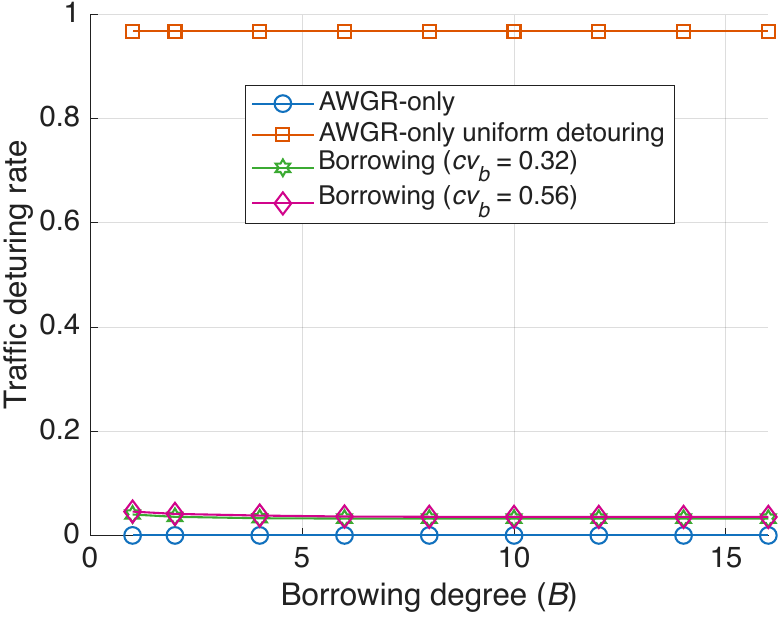}
        \caption{}
        \label{fig:detouring_gravity}
    \end{subfigure}
    \hfill
    \begin{subfigure}[b]{0.33\textwidth}
        \centering
        \includegraphics[height=4.4cm]{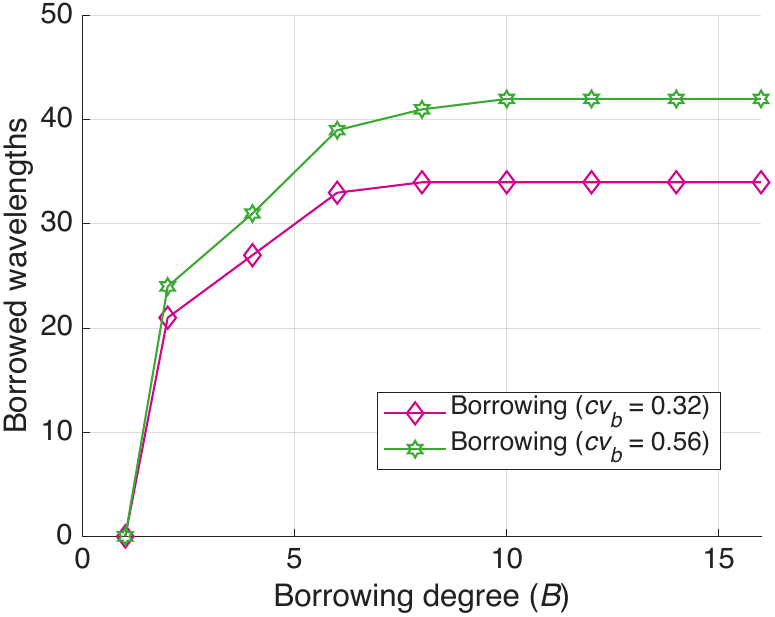}
        \caption{}
        \label{fig:borrowing_gravity}
    \end{subfigure}
    \hfill
    \begin{subfigure}[b]{0.33\textwidth}
        \centering
        \includegraphics[height=4.4cm]{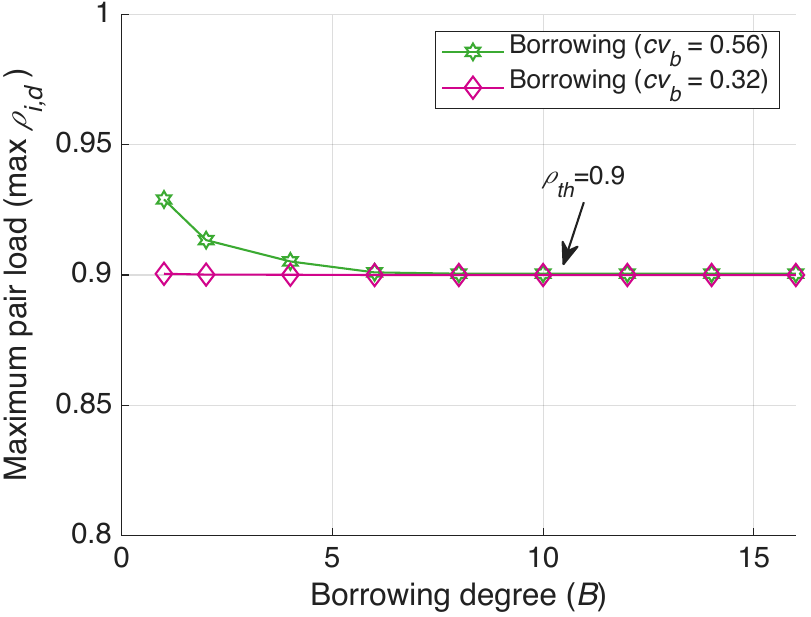}
        \caption{}
        \label{fig:load_gravity}
    \end{subfigure}
    \caption{Performance varying the borrowing degree $B$ with $\rho_\text{th}=0.9$ in a system with $W=32$ leaves and gravity-model traffic.}
    \label{fig:gravity}
\end{figure*}

\autoref{fig:gravity} shows the resulting performance. To compare it with the lognormal case, note that the coefficient of variation $cv$ of the off-diagonal entries $A^E_{i,j}$, measured over the generated matrices, is $0.439$ and $0.780$ for $cv_b=0.32$ and $cv_b=0.56$, respectively; these values are close to the $cv=0.5$ and $cv=1$ cases of the lognormal model.
The qualitative conclusions drawn for lognormal traffic in \autoref{fig:borrowing1} also hold with gravity-model traffic. In absolute terms, both the number of borrowed wavelengths and the detouring rate are lower than in the lognormal case, since the asymmetry of gravity-model matrices is smoother. As in the case of lognormal, for $B\geq 8$ performance no longer improves significantly and the target load $\rho_\text{th}=0.9$ is reached.

\section*{APPENDIX VI: Dynamic Reconfiguration Algorithm and Traffic Model}

Algorithms~\ref{alg:dyn} and~\ref{alg:mig} detail the algorithm discussed in \autoref{sec:dynamic}. In the plan stage, Algorithm~\ref{alg:greedy} also returns the list $\mathcal{H}$ of the borrowings it applied, in order of application. The reduction stage reverts the new borrowings of $\mathcal{H}$ from the last to the first until the number of new laser activations $N_\text{act} \leq V$. The number of lasers switched on when moving from $\mathbf{b}$ to $\mathbf{b}'$ is
\begin{align}\label{eq:nact}
N_\text{act}(\mathbf{b},\mathbf{b}')={}&\lvert\{(i,j,d): b_{i,j,d}=0,\ b'_{i,j,d}=1\}\rvert\nonumber\\
&+\lvert\{(j,d): \textstyle\sum_i b_{i,j,d}=1,\ \sum_i b'_{i,j,d}=0\}\rvert,
\end{align}
i.e., the new borrowings (first term) plus the lender re-activations (second term), where a lender regains its default wavelength. In the simulator code, this quantity is computed by the function \texttt{borrowing\_diff}.

\begin{algorithm}[!t]
\footnotesize
\caption{Dynamic reconfiguration: trigger, plan and reduction}
\label{alg:dyn}
\begin{algorithmic}[1]
\Require current $\mathbf{b}$, $B$, $\rho_\text{th}$, trigger factor $\tau$, transition limit $\rho^{\max}_\text{th}$, activation limit $V$
\Ensure $\mathbf{b}$, $\mathbf{w}$ for every traffic matrix
\ForAll{new traffic matrix $\mathbf{A}^E_k$}
    \State \# ---Try traffic-engineering-only---
    \State $\mathbf{w}, \boldsymbol{\rho}, \Delta\rho_{\max}, T^D \gets$ \textsc{2HWF}$(\mathbf{b})$ on $\mathbf{A}^E_k$
    \If{$\max_{i,d}\rho_{i,d} \le \tau\,\rho_\text{th}$} \Comment{reconfiguration trigger}
        \State \textbf{continue} \Comment{traffic-engineering is enough}
    \Else
        \State \# ---Plan phase---
        \State $\mathbf{b}^\text{plan}, \mathbf{w}^\text{plan}, \mathcal{H} \gets$ Algorithm~\ref{alg:greedy}$(\mathbf{A}^E_k, B, \rho_\text{th})$ 
        \State \# ---Reduction phase---
        \State $\mathbf{b}^\text{target} \gets \mathbf{b}^\text{plan}$ 
        \ForAll{$(i,j,d) \in \mathcal{H}$ with $b_{i,j,d}=0$} 
        \State \# new borrowing from the last to the first
            \If{$N_\text{act}(\mathbf{b},\mathbf{b}^\text{target}) \le V$} \textbf{break} \EndIf
            \State $\mathbf{b}' \gets \mathbf{b}^\text{target}$ with $b'_{i,j,d}=0$
            \If{$\exists\, i_0: b_{i_0,j,d}=1$} 
            \State \# $(j,d)$ was lent to an old borrower $i_0\neq i$
                \If{restoring $b'_{i_0,j,d}=1$ violates \eqref{eq:optc1}--\eqref{eq:optc5}} 
                    \State \textbf{continue} 
                \EndIf
                \State $b'_{i_0,j,d}=1$ \Comment{restoring attempt}
            \ElsIf{$\mathbf{b}'$ is not two-hop connected} 
                \State \textbf{continue}
            \EndIf
            \State $\mathbf{b}^\text{target} \gets \mathbf{b}'$ \Comment{restoring}
        \EndFor
        
        \If{$N_\text{act}(\mathbf{b},\mathbf{b}^\text{target}) > V$}
        \State \# Restoring old borrowings to decrease $N_\text{act}$ 
            \ForAll{$(i_0,j,d)$ with $b_{i_0,j,d}=1$, $\sum_k b^\text{target}_{k,j,d}=0$} 
                \State \# ${i_0,j,d}$ old borrowing not in $\mathbf{b}^\text{target}$
                \If{$N_\text{act}(\mathbf{b},\mathbf{b}^\text{target}) \le V$} \textbf{break} \EndIf
                \If{restoring $b^\text{target}_{i_0,j,d}=1$ satisfies \eqref{eq:optc1}--\eqref{eq:optc5}} 
                    \State $b^\text{target}_{i_0,j,d}=1$ 
                \EndIf
            \EndFor
        \EndIf
        \State \textsc{Migrate}$(\mathbf{b},\mathbf{b}^\text{target})$ \Comment{Algorithm~\ref{alg:mig}}
    \EndIf
\EndFor
\end{algorithmic}
\end{algorithm}

\begin{algorithm}[!t]
\footnotesize
\caption{Migration phase and transition test}
\label{alg:mig}
\begin{algorithmic}[1]
\Function{Migrate}{$\mathbf{b},\mathbf{b}^\text{target}$}
\State \# ---Migration phase---
\State $\mathcal{R} = \{(i,j,d): b_{i,j,d}=1,\ b^\text{target}_{i,j,d}=0\}$ \Comment{borrowing removals}
\State $\mathcal{A} = \{(i,j,d): b_{i,j,d}=0,\ b^\text{target}_{i,j,d}=1\}$ \Comment{new borrowings}
\Repeat 
    \State $\text{progress} = \textbf{false}$
    \ForAll{$(i,j,d) \in \mathcal{R}$}
        \State $\mathbf{c}^T \gets \mathbf{c}$ with $c^T_{i,d} = c_{i,d}-1$ 
        \State \# $i$ has lost the wavelength, $j$ has not regained it
        \State $\mathbf{w}^{Ti}, \boldsymbol{\rho}^T \gets$ \textsc{2HWF}$(\mathbf{c}^T)$ 
        \If{\textsc{TransOK}$(\mathbf{c}^T, \mathbf{w}^{Ti}, \boldsymbol{\rho}^T)$}
            \State \# start borrowing removal 
            \State apply $\mathbf{w}^{Ti}$ \Comment{traffic depletion}
            \State switch off at $i$
            \State switch on at $j$; wait until operative
            \State $b_{i,j,d} = 0$; $\mathbf{w}^{Tf}, \boldsymbol{\rho}, \Delta\rho_{\max}, T^D \gets$ \textsc{2HWF}$(\mathbf{b})$
            \State apply $\mathbf{w}^{Tf}$
            \State $\mathcal{R} \gets \mathcal{R} \setminus (i,j,d)$; $\text{progress} = \textbf{true}$
            \State \# removal completed
        \EndIf
    \EndFor
    \ForAll{$(i,j,d) \in \mathcal{A}$}
        \If{activating $b_{i,j,d}=1$ violates \eqref{eq:optc1}--\eqref{eq:optc5}}
        \State \textbf{continue}
        \EndIf
        \State $\mathbf{c}^T \gets \mathbf{c}$ with $c^T_{j,d} = 0$ 
        \State \# $j$ has given up the wavelength, $i$ has not received it
        \State $\mathbf{w}^{Ti}, \boldsymbol{\rho}^T \gets$ \textsc{2HWF}$(\mathbf{c}^T)$ \Comment{all traffic re-routed}
        \If{\textsc{TransOK}$(\mathbf{c}^T, \mathbf{w}^{Ti}, \boldsymbol{\rho}^T)$}
            \State \# start borrowing addition
            \State apply $\mathbf{w}^{Ti}$; \Comment{traffic depletion}
            \State switch off at $j$
            \State switch on at $i$; wait until operative
            \State $b_{i,j,d} = 1$; $\mathbf{w}^{Tf}, \boldsymbol{\rho}, \Delta\rho_{\max}, T^D \gets$ \textsc{2HWF}$(\mathbf{b})$
            \State apply $\mathbf{w}^{Tf}$
            \State $\mathcal{A} \gets \mathcal{A} \setminus (i,j,d)$; $\text{progress} = \textbf{true}$
            \State \# addition completed
        \EndIf
    \EndFor
\Until{\textbf{not} progress \textbf{or} $\mathcal{R} \cup \mathcal{A} = \emptyset$}
\EndFunction
\Statex
\Function{TransOK}{$\mathbf{c}^T, \mathbf{w}^{Ti}, \boldsymbol{\rho}^T$}
    \If{$\mathbf{c}^T$ is not two-hop connected} \Return \textbf{false} \EndIf
    \If{$\mathbf{w}^{Ti}$ assigns traffic to a pair with $c^T_{a,b}=0$} \Return \textbf{false} \EndIf
    \State \Return $\big[\rho^T_{a,b} \le \max(\rho^{\max}_\text{th}, \rho_{a,b})\ \ \forall (a,b)\big]$ \Comment{$\rho$: loads before the switch}
\EndFunction
\end{algorithmic}
\end{algorithm}

\subsubsection{Traffic model}
We use the same lognormal model of \autoref{sec:perf} with  mean $\overline{A^E}=0.65$, $cv=1$ and $W=32$. Time is slotted, with step index $k$. Every $\omega=96$ steps, i.e., at the \textit{renewal epochs} $k=n\omega$, $n=0,1,2,\dots$, a new matrix $\mathbf{A}^{E}(n\omega)$ is drawn independently from the model, so that the traffic at a renewal epoch is completely uncorrelated with that of the previous one; $\omega$ thus sets the long-term degree of variability of the traffic. In the intermediate steps $k=(n-1)\omega+1,\dots,n\omega-1$, a \textit{transitional} matrix $\mathbf{A}^E(k)$ moves smoothly from $\mathbf{A}^{E}((n-1)\omega)$ to $\mathbf{A}^{E}(n\omega)$, so that the traffic never stops moving. With $\alpha=(k-(n-1)\omega)/\omega$, the logarithm of the off-diagonal elements is obtained by interpolation, before the rescaling described below,
\begin{equation}
\ln \tilde A^{E}_{i,d}(k)=(1-\alpha)\ln A^{E}_{i,d}((n-1)\omega)+\alpha\ln A^{E}_{i,d}(n\omega),
\end{equation}
and its deviations from the mean are then rescaled so that their standard deviation is the interpolation of the two end-point standard deviations,
\begin{align}
&\ln A^{E}_{i,d}(k)=\mu(k)+\frac{\sigma(k)}{\tilde\sigma(k)}\left(\ln \tilde A^{E}_{i,d}(k)-\mu(k)\right)\\
&\sigma(k)=(1-\alpha)\,\sigma((n-1)\omega)+\alpha\,\sigma(n\omega),
\end{align}
where $\mu(k)$ and $\tilde\sigma(k)$ are the mean and the standard deviation of $\ln \tilde A^{E}_{i,d}(k)$ over the off-diagonal elements, and $\sigma(n\omega)$ is the standard deviation of $\ln A^{E}_{i,d}(n\omega)$ over the same elements. The interpolation is performed in the logarithmic domain because, there, a linear combination of the two end-point matrices is still normally distributed, so that after exponentiation the transitional matrices remain lognormal, and the elements remain positive. The rescaling then compensates for the variance reduction of the blend, so that the transitional matrices keep, approximately, the coefficient of variation $cv$ of the model instead of becoming smoother than the renewal matrices. 

\subsubsection{Sensitivity analysis versus maximum activation limit ($V$) and traffic renewal period ($\omega$)}

\begin{figure*}[!t]
    \centering
    \begin{subfigure}[b]{0.31\textwidth}
        \includegraphics[width=\textwidth]{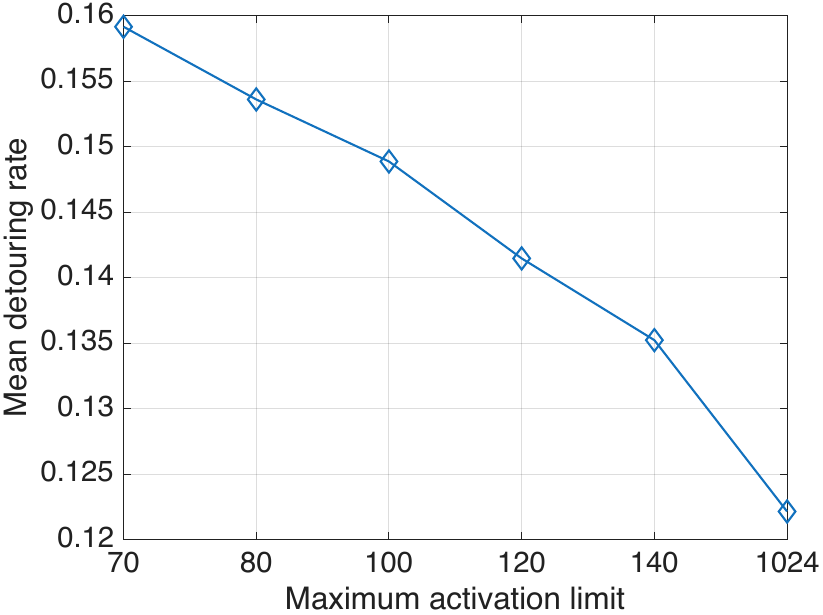}
        \caption{}
        \label{fig:dynamic_tv_vslimit}
    \end{subfigure}
    \hfill
    \begin{subfigure}[b]{0.31\textwidth}
        \includegraphics[width=\textwidth]{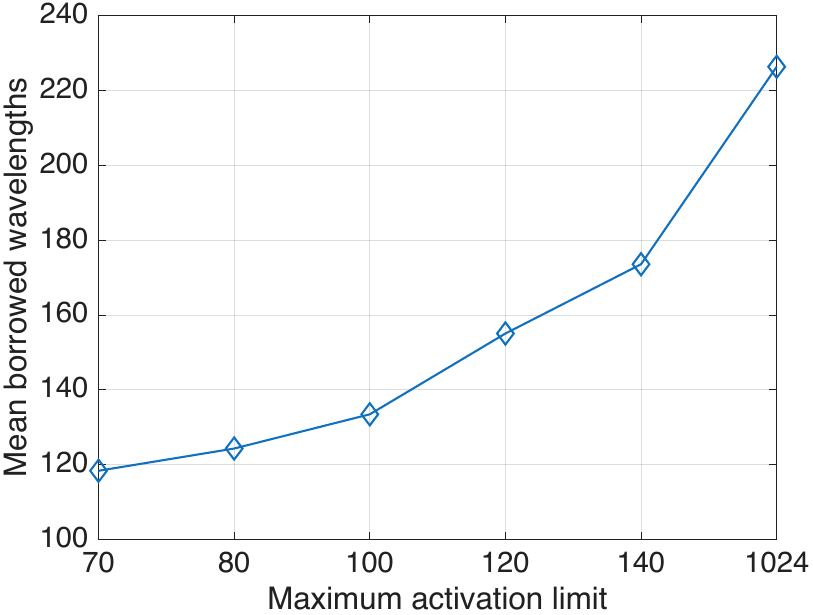}
        \caption{}
        \label{fig:dynamic_borrowing_vslimit}
    \end{subfigure}
    \hfill
    \begin{subfigure}[b]{0.31\textwidth}
        \includegraphics[width=\textwidth]{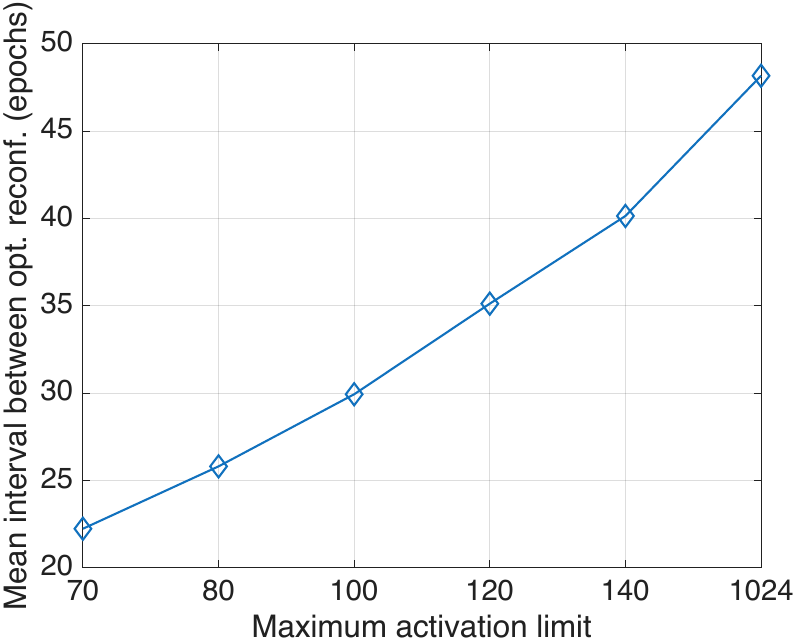}
        \caption{}
        \label{fig:dynamic_interval_vslimit}
    \end{subfigure}
    \caption{Dynamic regime with for a system with $W=32$ leaves, $B=8$ and $\omega=96$, $\rho_\text{th}=0.8$, $\tau=1.05$, $\rho^{\max}_\text{th}=0.95$, versus maximum activation limit ($V$)}
    \label{fig:activations}
\end{figure*}

\begin{figure*}[!t]
    \centering
    \begin{subfigure}[b]{0.31\textwidth}
        \includegraphics[width=\textwidth]{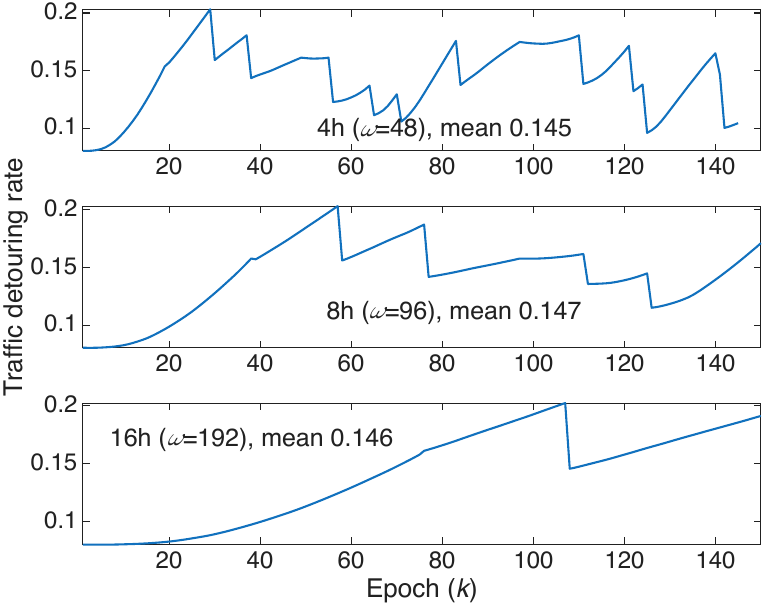}
        \caption{}
        \label{fig:dynamic_tv_vsperiod}
    \end{subfigure}
    \hfill
    \begin{subfigure}[b]{0.31\textwidth}
        \includegraphics[width=\textwidth]{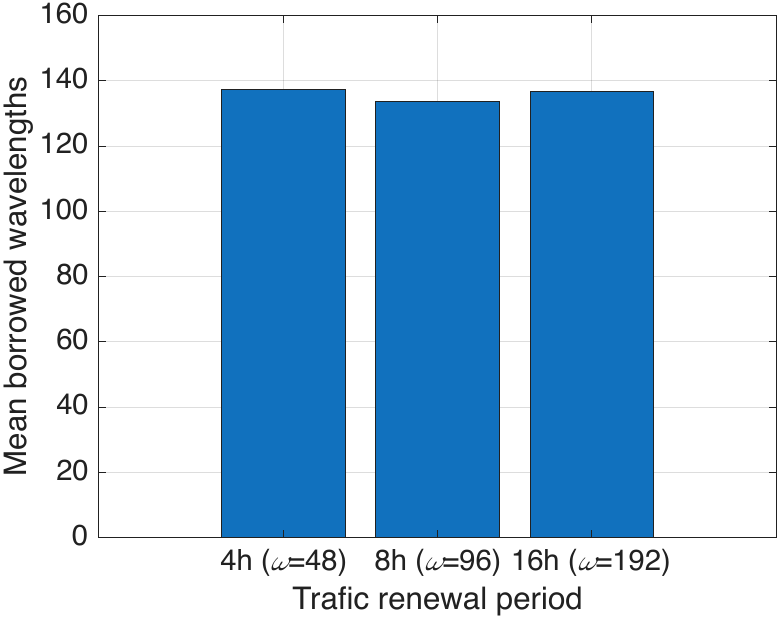}
        \caption{}
        \label{fig:dynamic_borrowing_vsperiod}
    \end{subfigure}
    \hfill
    \begin{subfigure}[b]{0.31\textwidth}
        \includegraphics[width=\textwidth]{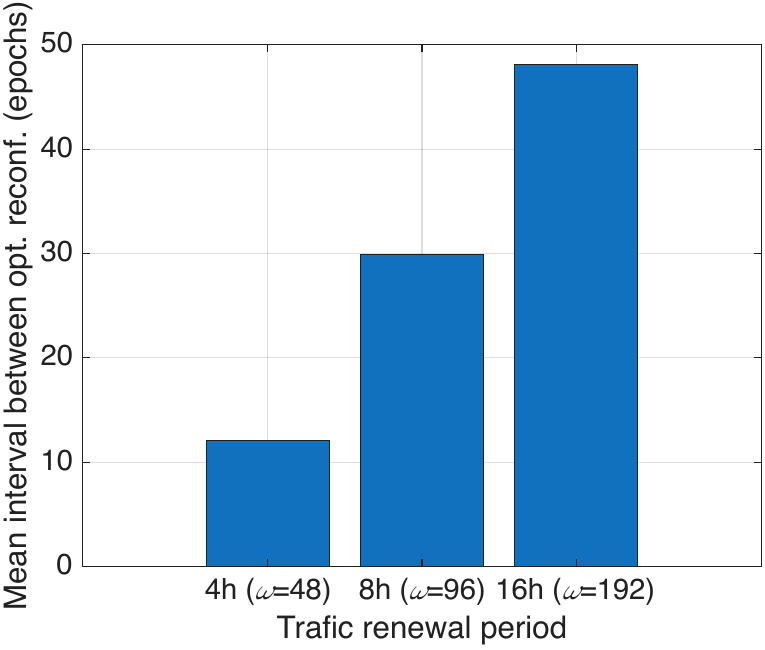}
        \caption{}
        \label{fig:dynamic_interval_vsperiod}
    \end{subfigure}
    \caption{Dynamic regime with for a system with $W=32$ leaves, $B=8$ and maximum activation limit $V=100$, $\rho_\text{th}=0.8$, $\tau=1.05$, $\rho^{\max}_\text{th}=0.95$, versus renewal period ($\omega$)}
    \label{fig:renewal}
\end{figure*}

This section shows a sensitivity analysis of the dynamic control policy discussed in \autoref{sec:dynamic} versus the maximum activation limit ($V$) and the traffic renewal period ($\omega$).

\autoref{fig:activations} shows the dynamic-regime performance versus the activation limit $V$ per migration, averaged over 288 control epochs. $V=1024=W^2$ represents the unlimited case, in which $\mathbf{b}^\text{target}=\mathbf{b}^\text{plan}$ of Algorithm~\ref{alg:greedy}. For all limited $V$ tested, every migration used the maximum number $V$ of activations, whereas in the unlimited case the average was 273.33.

As $V$ increases, the loop can exploit more reconfiguration possibilities: the detouring rate decreases, although only by a few percentage points, and the number of borrowings grows. A larger $V$ also yields configurations that better fit the underlying traffic matrix, which tolerate traffic variations longer before a new optical reconfiguration is needed, as confirmed by the longer interval between reconfigurations versus $V$. Conversely, a larger $V$ lengthens the migration, since more laser activations and link bring-ups are needed to complete it.

\autoref{fig:renewal} shows the dynamic-regime performance versus the traffic dynamicity, controlled by the parameter $\omega$, i.e. how many control epochs the traffic matrix takes to change completely. With an epoch of 5~min, the renewal periods considered are 4, 8 and 16 hours. The detouring rate and the maximum wavelength load simply stretch in time as the renewal period grows, and their average values, as well as the average number of borrowed wavelengths, are practically unchanged. A shorter renewal period instead reduces the time between migrations, since the optical topology must be reconfigured more frequently.

\section*{APPENDIX VII: Preliminary Power Budget}
\label{sec:power}
This appendix discusses a preliminary power budget of the architecture. For the AWG, we consider an insertion loss of 5~dB, counted twice since the end-to-end path traverses both an AWG mux and an AWG demux (10~dB total); for the OxC, 2~dB; and for the AWGR, we consider a loss varying with $W$, namely 4, 5, and 6~dB for $W=16,32,64$, respectively. For the combiner, we consider $10\log_{10}(B{=}8)$~dB, since a small $B$ (e.g., $B=8$) already achieves near-fully-reconfigurable performance (see \autoref{sec:perf})\footnote{Another possible implementation uses a WSS in combiner mode with near-zero combining loss, but requires active control coordinated with the OxC and incurs higher cost.}. We further consider a loss margin of 2~dB, accounting for connector loss and similar contributions. The resulting power loss is approximately 27, 28, and 29~dB for $W=16,32,64$, respectively.

Regarding the transmission power and receiver sensitivity of Ethernet transceivers, we consider those of the 400/800GBASE-DR8 \cite{ieee8023df}, namely 4~dBm for the TX power and $-5.9$~dBm for the sensitivity, yielding an approximate sustainable power loss of 10~dB. Consequently, DWDM-ready amplifiers in the 15--25~dB range, such as those provided by EDFAs, are required along the end-to-end path, e.g., after each coupler. 

The loss values above are based on the following sources. AWG: below 3.7~dB (40-ch 100~GHz) and up to 5.5~dB (80-ch 50~GHz, $W=64$) for commercial DWDM modules \url{https://edgeoptic.com}, \url{https://www.hilinktech.com/aawg/50ghz-aawg-dwdm-mux-demux-80ch.html}; we use 5~dB across all $W$. OxC: below 2~dB for 136$\times$136 MEMS switches \cite{poutievski2022jupiter,lightwave2023}. AWGR: consistent with \url{https://lumilaserchip.com/product-category/awgr/awgr-module/} and \cite{awgr64}. EDFA: commercial C-band pre-amplifiers reach up to 25~dB gain \url{https://www.optilab.com/products/c-band-pre-amp-edfa-module-14-dbm-25-db-gain}.

\end{document}